\documentclass[aps,twocolumn,superscriptaddress]{revtex4-1}

\usepackage{graphicx}
\usepackage{rotate}
\usepackage{epsfig}
\usepackage[dvipsnames]{xcolor}
\usepackage[normalem]{ulem}
\usepackage{amssymb,amsfonts,amsmath}
\usepackage{bm}
\usepackage{array}
\usepackage{mdwlist} 
\usepackage{subcaption}
\usepackage{mwe}
\usepackage{xr}

\usepackage{multibib}
\usepackage{colortbl}
\usepackage{url}
\usepackage{hyperref}

\definecolor{MyBrick}{rgb}{0.84,0.01,0.01}
\colorlet{MyBlue}{NavyBlue!95!white}

\newcommand{\etal}{\emph{et al.} }
\newcommand{\ie}{\emph{i.e.} }

\newcommand{\xxi}{x_i}
\newcommand{\xxj}{x_j}
\newcommand{\yyi}{y_i}
\newcommand{\yyj}{y_j}

\newcommand{\gloss}{GloSS }

\graphicspath{{pictures/}}

\begin{document}

\title{Irreducible network backbones:\\ unbiased graph filtering via maximum entropy}

\author{Valerio Gemmetto}
\email{gemmetto@lorentz.leidenuniv.nl}
\affiliation{Instituut-Lorentz for Theoretical Physics, Leiden Institute of Physics, University of Leiden, Niels Bohrweg 2, 2333 CA Leiden, The Netherlands}

\author{Alessio Cardillo}
\email{alessio.cardillo@epfl.ch}
\affiliation{Laboratory for Statistical Biophysics, \'Ecole Polytechnique F\'ed\'erale de Lausanne (EPFL), CH-1015 Lausanne, Switzerland}
\affiliation{Institute for Biocomputation and Physics of Complex Systems (BIFI), University of Zaragoza, E-50018 Zaragoza, Spain}

\author{Diego Garlaschelli}
\email{garlaschelli@lorentz.leidenuniv.nl}
\affiliation{Instituut-Lorentz for Theoretical Physics, Leiden Institute of Physics, University of Leiden, Niels Bohrweg 2, 2333 CA Leiden, The Netherlands}


\begin{abstract}
Networks provide an informative, yet non-redundant description of complex systems only if links represent truly dyadic relationships that cannot be directly traced back to node-specific properties such as size, importance, or coordinates in some embedding space. 
In any real-world network, some links may be reducible, and others irreducible, to such local properties.
This dichotomy persists despite the steady increase in data availability and resolution, which actually determines an even stronger need for filtering techniques aimed at discerning essential links from non-essential ones.
Here we introduce a rigorous method that, for any desired level of statistical significance, outputs the network backbone that is irreducible to the local properties of nodes, \ie their degrees and strengths. 
Unlike previous approaches, our method employs an exact maximum-entropy formulation guaranteeing that the filtered network encodes only the links that cannot be inferred from local information.
Extensive empirical analysis confirms that this approach uncovers essential backbones that are otherwise hidden amidst many redundant relationships and inaccessible to other methods. 
For instance, we retrieve the hub-and-spoke skeleton of the US airport network and many specialised patterns of international trade. 
Being irreducible to local transportation and economic constraints of supply and demand, these backbones single out genuinely higher-order wiring principles.
\end{abstract}

\maketitle
\section{Introduction}

Over the last two decades, networks have become a standard tool of analysis of many complex systems. 
A network can represent any instance of a natural, social or technological system as a collection of \emph{nodes} (or \emph{vertices}) connected by \emph{links} (or \emph{edges})~\cite{boccaletti-physrep-2006, newman-book-2010, barabasi-natphys-2011,liljeros-nature-2001, bullmore-nat_rev_neuro-2009, maslov-science-2002, guimera-science-2005, barthelemy-physrep-2011}. In many cases, it is possible to quantify the magnitude of interaction between two nodes, and encode it as a numerical value -- the \emph{weight} -- attached to the link connecting them. In these cases, we refer to those networks as \emph{weighted}~\cite{barrat-pnas-2004}. Weighted networks are the focus of this paper.

A recurrent and robust feature observed in the vast majority of real-world networks is a striking heterogeneity of the topological properties of different nodes. 
For instance, both the number of connections of a node (the so-called \emph{degree}) and the total weight of these connections (the so-called \emph{strength}) differ greatly across nodes in a network. 
Typically, the empirical distribution of both quantities in a given network is a power law, or more generally a broad distribution with `fat tails'.
Various models have been introduced in order to propose explanations for the origin of such distributions in real networks.
Regardless of the possible mechanisms generating it, the observed heterogeneity of nodes has immediate consequences for the way real networks can or should be analysed. 
For instance, topological quantities averaged or summed over nodes (such as the total number of links, or the total link weight) are typically not informative about the whole network structure. 
Consequently, many structural quantities (such as the \emph{clustering coefficient} measuring the relative abundance of triangles) should be defined \emph{locally} and interpreted conditionally on the values of the degree and/or strength of nodes. 

This paper focuses on another important and well known consequence of the heterogeneity of nodes: namely, the impossibility of using a single reference value, or equivalently a unique global threshold, to assess the importance of different links in a given weighted network.
If nodes have different strengths, global thresholds do not work, due to varying levels of statistical significance: a light-weighted link connecting two nodes with low strength may be even more significant than a heavy one connecting nodes with high strength~\cite{granovetter-ajsoc-1973}.
This calls for techiques to assess the statistical significance of links based on the local properties of nodes.
This problem arises in a variety of circumstances.
One of the most important and recurrent examples is \emph{graph filtering}, \ie the identification of the most relevant links  and the subsequent elimination of the least significant ones.
Of course, whether a link qualifies as `relevant' is largely problem-dependent and ultimately relates to the specific reason why the network is being filtered in the first place.
Such a reason may be of practical or fundamental character. In this paper, we propose a novel method of graph filtering whose motivation encompasses both aspects.

At a practical level, a widespread reason for filtering is the necessity of keeping real graphs \emph{sparse} in order to directly and easily pinpoint the main relationships between the units of the system they represent.
The recent data deluge associated with the so-called \emph{Big Data} era \cite{lynch-nature-2008} has bolstered the tendency to represent complex systems as networks \cite{onnela-pnas-2007, blondel-epjds-2015, menche-science-2015, myers-www-2014}. Unfortunately, the huge opportunities associated to the availability of big data are not free of charges. 
The increase of information, in fact, usually produces a raise in the number of reported connections, which in turn increases the computational complexity of most algorithms used for network analysis and visualization. 
Effectively, this makes real networks harder to render, characterize and ultimately get a grip on. 
To continue using networks to retrieve the essential \emph{backbone} of a complex system, uncluttering techniques are therefore needed.

At a more fundamental level, and irrespective of computational aspects, there is a subtler and less contemplated, yet very important reason for graph filtering.
While this is seldom acknowledged, the network representation of a real system is non-redundant, thus really necessary, only if the presence and/or magnitude of the pairwise interactions between the units of the system cannot be entirely inferred from node-specific properties -- such as size, importance, position in some underlying geometry, etc. 
Indeed, if node-specific properties were enough to characterize, infer or reconstruct the relationships among pairs of nodes (for instance if the network were a regular lattice embedded in space and the coordinates of nodes were given, or if the topology were a function of only the sizes of nodes and such sizes were known), then the network representation, while still correct, would be redundant. 
In general, some of the links of a given network may be `reducible', and others `irreducible', to node-specific properties.
If so, the former would be in some sense unsurprising, while the latter would be much more interesting and provide truly dyadic information.
This possibility calls for the introduction of \emph{graph filtering techniques that allow irreducible links to be discerned from reducible ones, thus highlighting the non-redundant backbone of a network.} 
One should at this point note that recently, motivated by the fact that certain (e.g. financial) networks cannot be empirical observed in their entirety because of privacy or confidentiality reasons, there has been a proliferation of techniques devised for reconstructing the hidden topology of such networks from partial information~\cite{review,book}. 
A specific class of methods have significantly increased the level of predictability of network structure from local, node-specific information~\cite{mastrandrea-njp-2014,mastrandrea-pre-2014,cimini-pre-2015,cimini-2015-scirep,squartini-2017-ans}, thus showing that, indeed, real networks can have a considerably big `reducible' component.

Taken together, the above two considerations imply that, as the empirical availability of node-related information increases and the toolkit for reconstructing networks from partial information expands, the question of what makes up the non-redundant, genuinely dyadic properties of a network becomes more important and more difficult to answer.
In this paper we focus on the problem of identifying the \emph{irreducible backbone} of a weighted network from a rigorous standpoint.
We define such backbone as the collection of links that cannot be reconstructed via unbiased inference from the knowledge of the local topological properties of nodes.
This definition guarantees that, by construction, the network backbone only encodes truly dyadic information.
We identify the irreducible backbone by constructing, for a given empirical network, a corresponding unbiased maximum-entropy null model where both the degree and the strength of each node are preserved as ensemble averages. 
Such maximum-entropy model is described by the generalized Bose-Fermi distribution~\cite{bose-fermi} and is called the \emph{Enhanced Configuration Model} (ECM)~\cite{mastrandrea-njp-2014} because it enhances the Weighted Configuration Model (\ie the maximum-entropy ensemble of weighted networks with given node strengths) by adding the degree sequence as an extra constraint.

The choice of the ECM as our null model is motivated by a recent series of theoretical and empirical results focusing on the problem of network reconstruction~\cite{mastrandrea-njp-2014,mastrandrea-pre-2014,cimini-pre-2015,cimini-2015-scirep,squartini-2017-ans}. These studies have shown that the information encoded in the degrees and strengths of the nodes can be used to replicate many higher-order properties of real-world networks~\cite{mastrandrea-njp-2014,mastrandrea-pre-2014}. 
The knowledge of both degrees and strengths is crucial to this purpose.
Indeed, using only the degrees (as in the Binary Configuration Model~\cite{randomizing1,randomizing3}) would provide no information about link weights, while using only the strengths (as in the Weighted Configuration Model~\cite{randomizing2,randomizing3}) would lead to an exceedingly high density of links resulting, tipically, in almost complete graphs.
Consequently, when the degrees of nodes are not empirically accessible, the success of the reconstruction method entirely depends on how well one is able to preliminarily obtain reliable estimates of the degrees, before constructing maximum-entropy null models that simultaneously preserve the (observed) strengths and the (inferred) degrees~\cite{mastrandrea-pre-2014,cimini-pre-2015,cimini-2015-scirep,squartini-2017-ans}.
In conclusion, these studies show that, in presence of only local node-specific information, the best unbiased inference about the entire structure of a weighted network is the one obtained using  both the strengths and the degrees of all nodes as input.

The rest of the paper is organized as follows.
In sec.~\ref{sec:previous} we discuss the novel ingredients of our framework with respect to previous graph filtering approaches.
In sec.~\ref{sec:model} we describe our method for the extraction of irreducible network backbones in detail.
In sec.~\ref{sec:results} we show the results of an extensive empirical analysis using our method.
In sec.~\ref{sec:specifications} we discuss further extensions of our method to directed and bipartite networks.
Finally, in sec.~\ref{sec:conclusions} we make some concluding remarks.





\section{Relation to previous work\label{sec:previous}}

Despite our motivation, \ie the identification of the irreducible network backbone that cannot be inferred from the local topological properties, is quite novel, our work necessarily relates to previous literature.
Under different names (e.g. thresholding, filtering, pruning, sparsification, backbone extraction, statistical validation, etc.), several algorithms have been proposed in order to remove links from a network. 
In general, the available approaches can be grouped in two main categories: \emph{coarse graining} and \emph{edge removal} methods. 
The former tend to merge together nodes with similar properties, thereby providing a hierarchical, multiscale view of the system \cite{kim-prl-2004, gfeller-prl-2007}. 
The latter fix instead the scale and proceed by removing connections. 
Edge removal methods split further into two sub-categories: \emph{pruning} and \emph{sparsification} techniques. Pruning approaches aim at removing connections to unveil some hidden structure/property of the system that is considered to be unknown \emph{a priori}. On the other hand, sparsification techniques remove connections while preserving some property of the original system, thus aiming at retrieving comparable information but at a cheaper computational cost \cite{hamann-arxiv-2016,coscia-arxiv-2017,sparsification}.

Among the pruning solutions, the most straightforward one is
thresholding and its variations \cite{de_vico_fallani-pcompbio-2017}. Removing all the edges having a weight, $w$, lighter than a given value, $w_t$, produces systems surely sparser but at the cost of losing all their ``weak ties'' \cite{granovetter-ajsoc-1973} and, more importantly, losing the weight heterogeneity which represents one of the hallmarks of complex systems \cite{barrat-pnas-2004}. Despite such serious limitations, thresholding has been used extensively, for example, in brain networks \cite{bullmore-nat_rev_neuro-2009}. Another notorious technique is the extraction of the Minimum Spanning Tree (MST) albeit it delivers an over simplification of the system because it destroys many features (cycles, clustering and so on) \cite{kruskal-pamsoc-1956, macdonald-epl-2005, scellato-epjb-2006}. 
Other pruning techniques are \emph{link validation} methods, which produce what is sometimes called a \emph{statistically validated network}~\cite{tumminello-pone-2011}.
These are the most similar to our method proposed here, yet still different, because in general the statistically validated network is not guaranteed to be irreducible (e.g. if the full information about strengths and degrees is not specified) or unbiased (e.g. if the procedure is not maximum-entropy).
To the best of our knowledge, in fact, validation of empirical networks against maximum-entropy ensembles with constrained strenghts and degrees has not be done yet.

On the other hand, the idea of sparsification relies on the (often implicit) assumption that the properties of the original network, especially those to be preserved, are statistically significant and therefore worth preserving in the first place.
This means that sparsification techniques require, at least in principle, that a preliminary filtering has already taken place.
In this sense, the method we propose in this paper is a filtering method that can be used for link pruning, link validation, and as a preliminary step for other sparsification methods.


Coming to a closer inspection of the available techniques, three methods are most directly related to what we are going to develop in this paper: the Disparity Filter (DF) introduced by Serrano \etal \cite{serrano-pnas-2009}, the so-called \gloss method proposed by Radicchi \etal \cite{radicchi-pre-2011}, and a method recently developed by Dianati \cite{dianati-pre-2016}.

\subsection{The Disparity Filter}
The DF is very close in spirit to our method, since it constrains both the strength and the degree of each node~\cite{serrano-pnas-2009}. However, there are crucial differences that we now explain. 

The DF assumes that, in the null model, the strength $s_i$ of each node $i$ is redistributed uniformly at random over the $k_i$ links of that node.
The resulting criterion for establishing whether a link having weight $w_{ij}^{\ast}$ satisfies the null hypothesis requires the computation of the following $p$-value
\begin{equation}
\gamma_{ij} = 1 - \int_{0}^{w_{ij}^{\ast}} \rho(w|s_i,k_i) \, dw \,,
\label{eq:serrano}
\end{equation}
and its comparison with a chosen critical value $\widetilde{\gamma}$.
Here $\rho(w|s_i,k_i)$ is probability (density) that a link has weight $w$, under the null hypothesis that the value $s_i$ is partitioned uniformly at random into $k_i$ terms. 
By contrast, we will see that the correct maximum-entropy probability derived in our approach does not correspond to such uniformly random partitioning, as it collectively depends also on the strengths and degrees of all other nodes. This means that the DF introduces some bias. 
This bias can be understood by noticing that, when distributing the strength of a node at random over its links, it disregards the strength of the nodes at the other end of these links, thus effectively `flattening' the weights received by these nodes.
For each node $i$, this creates randomized weights that tend to be too small around high-strength neighbours and too small around low-strength neighbours.
As a consequence, as we will confirm later, the DF has a bias towards retaining heavier connections.

From a mathematical point of view, the above problem is manifest in the fact that, despite $w^\ast_{ij}=w^\ast_{ji}$ (we are considering undirected networks for the moment), eq.~\eqref{eq:serrano} is not symmetric under the exchange of $i$ and $j$, and in general one has $\gamma_{ij}\ne\gamma_{ji}$.
To partly compensate for this, the DF is usually applied twice, from the perspective of each of the two endpoints of an edge.
However, it should be noted that, in general, the resulting randomized weights cannot be actually generated in any network, as it is not possible to produce randomized link weights that realize the null hypothesis for all nodes simultaneously.
So, in the end, the statistical test is based on an ill-defined null hypothesis.

Another evidence of the above problem is the fact that, under the correct maximum-entropy model, the level of heterogeneity of the weights of the links incident on a node (as measured for instance by the so-called `disparity') does not take on the values one usually expects under the assumption of uniform randomness and is strongly biased towards other values~\cite{bose-fermi}.
A consequent bias must necessarily arise in the $p$-values as calculated above.

A final difference with respect to our method is that the DF enforces the degrees and strengths sharply (i.e. as in the \emph{microcanonical} ensemble in statistical physics), whereas we enforce them only as ensemble averages (i.e. as in the \emph{canonical} ensemble)~\cite{squartini-njp-2015}. 
The microcanonical implementation, even if carried out in a correct and unbiased way, implies statistical dependencies between the weights of all edges in the null model, because these weights must add up to a deterministic value. This in turn implies that the statistical test cannot be carried out separately for each edge.
In our canonical implementation of the null model, all edges are instead independent, a property that allows us to consistently carry out the statistical test for each node separately, even if, as we mentioned, our $p$-value for each link depends on the degrees and strengths of all other nodes in the network, as desired.
The recent results about the non-equivalence of microcanonical and canonical ensembles of random graphs with given strengths and/or degrees~\cite{squartini-njp-2015,squartini-prl-2015,garlaschelli-jpa-2016} imply that the two approaches remain different even in the limit of large network size, and must therefore lead to different results.

\subsection{The \gloss method}
In the \gloss method, the null model used to assign $p$-values to edges is a network with exactly the same topology of the original network and with link weights randomly drawn from the empirical weight distribution $P(w)$. 
This effectively means that the observed link weights are randomly reshuffled over the existing, fixed topology.
Unlike the local criterion of the DF, this choice results in a \emph{global} null model.
Indeed, since links are fixed and they all have the same probability of being assigned a given weight, the statistical test is the same for every existing edge, and the method effectively reduces to selecting the strongest weight only, thus setting a global threshold which depends on the desired confidence level. This leads us back to the problem of global thresholds being inappropriate for networks with strong heterogeneity.

Another, related problem with \gloss is the fact that it conceives the topology and the weights as two separate, or separable, network properties. This is however hard to justify, since the topology is encoded in the adjacency matrix whose entries $\{a_{ij}\}=\{\Theta(w_{ij})\}$ are binary projections of the link weights $\{w_{ij}\}$, and thus entirely dependent on the latter. 
Indeed, as a result of this decoupling, the null model turns light links into heavy ones and viceversa, irrespective of the importance of the end-point vertices. In other words, it views the weight distribution as unconditional on the strengths of the end-point vertices. The strengths are viewed as the result of, rather than a constraints for, a random realization of weights.
The resulting expected strengths are indeed proportional to the observed degrees.
One can partly relax the null model by globally reshuffling the weights while simultaneously randomizing the topology in a degree-preserving way~\cite{ramasco}, but the method will still retain the proportionality between the expected strength and the degree of a node, and the underlying notion of complete separability of topology and weights.

It should be noted that the proportionality between strengths and degrees in the null model violates the strongly non-linear relationship observed between these two quantities in real-world networks~\cite{barrat-pnas-2004,cimini-pre-2015,cimini-2015-scirep}. 
Coming back to the network reconstruction problem, this implies that \gloss suffers from what we may call a \emph{redundancy} problem: since many properties of real-world networks can be inferred from the empirical degrees and strengths of nodes, \gloss cannot ensure that the filtered network is irreducible to the knowledge of such node-specific properties.
Indeed, such properties contain in general more information than what is retained in the null model. 
The resulting network backbone may therefore still contain redundant interactions.

\subsection{The `hairball' method}
Dianati has recently proposed a different `hairball' approach where, unlike the methods discussed above, the null model is maximum-entropy based and therefore unbiased. The constraints imposed on entropy maximization are the strenghts of all nodes, but not the degrees. Dianati considers two distinct null models: a local one, acting on single links, named Marginal Likelihood Filter (MLF) and a global one, acting on the network as a whole, named Global Likelihood Filter (GLF). Both null models produce graphs which, for a given $p$-value, are quite alike.

Although the maximum-entropy nature of the filter introduced by Dianati fixes the problem of bias encountered by the other approaches, the method still suffers from the redundancy problem, even if in a direction in some sense opposite to that of GloSS. 
Concretely, constraining only the strength sequence in the null model corresponds to generating almost complete networks~\cite{randomizing2,randomizing3}. 
This implies that the nonlinear empirical strength-degree relation is again violated, here because the degree of each node tends to saturate to the maximum allowed value and is therefore independent of the strength. 
Once more, this does not guarantee that the filtered network is irreducible to the knowledge of the degrees and strenghts of nodes.
Indeed, the weights are redistributed among virtually all the possible pairs of nodes, which also means that the empirical non-zero link weights are systematically larger than those generated by the null model.
This implies that the filter tends to retain too many spurious links.

\section{Extraction of irreducible backbones: the ECM filter\label{sec:model}}

In the rest of this paper, we aim at combining together the good ingredients of previous methods, while overcoming their most important limitations.
In particular, we want to retain the unbiasedness of the maximum-entropy approach proposed by Dianati while retaining the empirical non-linear relationship between strengths and degrees as in the DF.

We therefore introduce a new filtering method based on the comparison between a given real-world weighted network and a canonical \emph{maximum-entropy} ensemble of weighted networks having (on average) the same degree sequence and the same strength sequence as the real network, \ie the ECM.
Our model can be fully characterized analytically, a property that allows us to explicitly calculate the exact $p$-value for each realized edge in the original network.
Unlike the DF, our maximum-entropy construction ensures consistency from the point of view of both nodes at the endpoint of an edge and makes the null hypothesis realizable by the networks in the statistical ensemble.
It also makes different edges statistically independent, thus justifying the establishment of a separate test for each observed link.
We remark that, unlike all previous approaches, the use of the ECM ensures that the filtered network cannot be retrieved by any impartial and unbiased network reconstruction method that starts from local information. This also fixes the redundancy problem of other methods.

Below, we first briefly review the definition and main properties of the ECM~\cite{mastrandrea-njp-2014,squartini-njp-2015}, which has been originally introduced under the name of \emph{Bose-Fermi} ensemble~\cite{bose-fermi}, and then provide a new recipe to use it for the purpose of graph filtering. 

\subsection{The Enhanced Configuration Model or Bose-Fermi Ensemble}
Very generally, a maximum-entropy model is a canonical ensemble described by a probability distribution $P(G)$ (over the microscopic configurations $\{G\}$ of the system) that maximizes the Shannon-Gibbs entropy $S=-\sum_{G}P(G)\ln P(G)$, while satisfying a given set of macroscopic constraints enforced as ensemble averages \cite{park-pre-2004}. The formal solution to this problem is a Boltzmann-like (\ie exponential) probability function of the form $P(G)=Z^{-1}e^{-H(G)}$, whose negative exponent, sometimes (improperly) termed ``Hamiltonian (function)'' $H(G)$, is a linear combination of the constraints and whose normalization constant is the inverse of the so-called ``partition function'' $Z=\sum_G e^{-H(G)}$. We are now going to define these quantities rigorously for the case of interest to us.

We consider an ensemble ${\cal W}$ of undirected, unipartite weighted networks with a fixed number $N$ of nodes (an explicit generalization to the case of directed and bipartite networks is provided later in sec.~\ref{sec:specifications}).
Each element of $\cal{W}$ is a weighted graph, uniquely specified by a $N\times N$ symmetric matrix $\bm{W}$ whose entry $w_{ij}=w_{ji}$ represents the weight of the link connecting node $i$ to node $j$ ($w_{ij}=0$ means that $i$ and $j$ are not connected). Without loss of generality, we assume integer weights ($w_{ij}=0,1,2,\dots$) and no self-loops ($w_{ii}=0$ for all $i$). 
Starting from the matrix $\bm{W}$, one can construct the \emph{adjacency matrix} $\bm{A}(\bm{W})$ whose entry is defined as $a_{ij}(\bm{W})=\Theta(w_{ij})$, \ie $a_{ij}(\bm{W})=1$ if $w_{ij}>0$ and $a_{ij}(\bm{W})=0$ if $w_{ij}=0$.
Given a network $\bm{W}$, the \emph{strength} of node $i$ is defined as $s_i(\bm{W})=\sum_{j\ne i} w_{ij}$ and the \emph{degree} of node $i$ is defined as $k_i(\bm{W})=\sum_{j\ne i}a_{ij}(\bm{W})$.

Let us consider an empirical network, $\bm{W}^{\ast}$, that we would like to filter.
We define $s^{\ast}_i\equiv s_i(\bm{W}^{\ast})$ and $k^{\ast}_i\equiv k_i(\bm{W}^{\ast})$, so that the resulting \emph{empirical strength and degree sequences} are denoted as $\vec{k}^{\ast}$ and $\vec{s}^{\ast}$ respectively.
We look for the probability distribution $P$ over graphs that maximizes the entropy, under the constraint that the expected degree and strength of each node equal the empirical values, e.g. 
\begin{equation}
\langle \vec{k}\rangle=\vec{k}^{\ast},\quad \langle \vec{s}\rangle=\vec{s}^{\ast}.
\label{eq:constraints}
\end{equation}
The above requirement introduces a Lagrange multiplier, which for later convenience we denote as $-\ln x_i$ (with $x_i>0$ for all $i$), for each expected degree $\langle k_i\rangle$ and another multiplier, denoted as $-\ln y_i$ (with $0<y_i<1$ for all $i$), for each expected strength $\langle s_i\rangle$. 
The graph probability we are looking for will depend on these $2N$ parameters, which we array in two $N$-dimensional vectors $\vec{x}$ and $\vec{y}$.
We require that such probability, denoted as  $P(\bm{W} | \vec{x},\vec{y})$ from now on, maximizes the Shannon-Gibbs entropy 
\begin{equation}
S(\vec{x},\vec{y}) = - \sum_{\bm{W}\in \cal{W}}^{} P(\bm{W} | \vec{x},\vec{y}) \, \ln P(\bm{W} | \vec{x},\vec{y}) \,
\label{eq:entropy}
\end{equation}
subject to the constraints in~\eqref{eq:constraints} and to the normalization condition
\begin{equation}
\sum_{{\bm W}\in\cal{W}}P(\bm{W} | \vec{x},\vec{y})=1.
\end{equation}
The solution to the above constrained maximization problem is found to be~\cite{bose-fermi,mastrandrea-njp-2014,squartini-njp-2015} the probability
\begin{equation}
P(\bm{W}|\vec{x},\vec{y}) = \dfrac{e^{-H \left( \bm{W}|\vec{x},\vec{y} \right)}}{Z(\vec{x},\vec{y})}=\prod_{i=1}^N \prod_{j<i} q_{ij}(w_{ij}),
\label{eq:graph_prob}
\end{equation}
where we have introduced the Hamiltonian
\begin{eqnarray}
H \left( \bm{W} \right|\vec{x},\vec{y})& =& -\sum_{i=1}^{N}\left[ k_i\left( \bm{W} \right)\ln x_i + s_i \left( \bm{W} \right)\ln y_i\right]\\
& =& -\sum_{i=1}^N\sum_{j<i} \left[ \Theta(w_{ij})\ln(x_i x_j)+  w_{ij}\ln(y_iy_j) \right],\nonumber
\end{eqnarray}
the partition function
\begin{eqnarray}
Z(\vec{x},\vec{y}) &=& \sum_{\bm{W}\in \cal{W}} e^{-H \left( \bm{W}|\vec{x},\vec{y} \right)}\\
 &=& \prod_{i=1}^N\prod_{j<i}\dfrac{1 - y_i y_j + x_i x_j y_i y_j}{1 - y_i y_j},\nonumber
\end{eqnarray}
and the probability that a link between nodes $i$ and $j$ has weight $w$:
\begin{eqnarray}
q_{ij}(w)&\equiv&\dfrac{\left( \xxi \xxj \right)^{\Theta(w)} \left( \yyi \yyj \right)^{w}  \left( 1 - \yyi \yyj \right) }{1 - \yyi \yyj + \xxi \xxj \yyi \yyj}\label{eq:qij}\\
&=&\left\{\begin{array}{ll}1-p_{ij}&\textrm{ if}\quad w=0\\
p_{ij}\left(\yyi\yyj\right)^{w-1}(1-\yyi\yyj)&\textrm{ if}\quad w>0\end{array}\right.,\nonumber
\end{eqnarray}
with 
\begin{equation}
p_{ij}\equiv1-q_{ij}(0)=\dfrac{\xxi \xxj \yyi \yyj}{1 - \yyi \yyj + \xxi \xxj \yyi \yyj}
\label{eq:pij}
\end{equation}
representing the probability that nodes $i$ and $j$ are connected, irrespective of the weight $w_{ij}>0$ of the link connecting them.
%

The key quantity describing the above maximum-entropy ensemble is $q_{ij}(w)$, whose expression~\eqref{eq:qij} has been first derived in~\cite{bose-fermi} and denoted as \emph{Bose-Fermi} distribution. The name comes from the fact that, as a result of enforcing both degrees and strenghts, the distribution combines features of the Bose-Einstein distribution, which is encountered when dealing with systems described by integer configurations such as $\bm{W}$, and the Fermi-Dirac distribution, which is encountered when dealing with systems described by binary configurations such as $\bm{A}(\bm{W})$.

We now come back to the real-world network ${\bm W}^{\ast}$ that we want to prune, and to its strength and degree sequences $\vec{s}^{\ast}$ and $\vec{k}^\ast$. 
Using the above form of $q_{ij}(w)$, $\langle \vec{s}\rangle$ and $\langle \vec{k}\rangle$ can be calculated explicity, both as functions of $\vec{x}$ and $\vec{y}$, so that the condition~\eqref{eq:constraints} can be rewritten explicitly~\cite{mastrandrea-njp-2014,squartini-njp-2015} as
\begin{eqnarray}
k^{\ast}_i&=&\sum_{j\ne i}\frac{\xxi \xxj \yyi \yyj}{1 - \yyi \yyj + \xxi \xxj \yyi \yyj}\quad\forall i,\label{eq:k}\\
s_i^{\ast}&=&\sum_{j\ne i}\frac{\xxi \xxj \yyi \yyj}{(1 - \yyi \yyj + \xxi \xxj \yyi \yyj)(1-y_iy_j)}\quad\forall i.\label{eq:s}
\end{eqnarray}
The above system of $2N$ coupled nonlinear equations is solved by certain parameter values $(\vec{x}^{\ast},\vec{y}^{\ast})$. 
Equivalently, the values $(\vec{x}^{\ast},\vec{y}^{\ast})$ can be proven to coincide with the values that maximize the log-likelihood of the model ~\cite{mymaxlikelihood,squartini-njp-2011,squartini-njp-2015}, \ie
\begin{equation}
(\vec{x}^{\,\ast}, \vec{y}^{\,\ast})=\underset{\{x_i>0,\, 0<y_i<1\,\forall i\}}{\mathrm{argmax}}\,\mathcal{L}(\vec{x},\vec{y})
\label{eq:ML}
\end{equation}
where the log-likelihood, $\mathcal{L}$, is defined as
\begin{eqnarray}
\mathcal{L}(\vec{x},\vec{y}) &=&  \ln P(\bm{W^{\ast}} | \vec{x},\vec{y}) \,\label{eq:LL}\\
&=&\sum_{i=1}^N \sum_{j<i} \ln q_{ij}(w^{\ast}_{ij})\nonumber\\
&=&\sum_{i=1}^{N}\left[ k_i^\ast\ln x_i + s_i ^\ast\ln y_i\right]\nonumber\\
&&-\sum_{i=1}^N\sum_{j<i}\ln \dfrac{1 - y_i y_j + x_i x_j y_i y_j}{1 - y_i y_j}.\nonumber
\end{eqnarray}

Once $P(\bm{W}|\vec{x},\vec{y})$ is evaluated at the parameter values $(\vec{x}^{\ast},\vec{y}^{\ast})$, we obtain the explicit maximum-entropy probability distribution $P(\bm{W}|\vec{x}^{\ast},\vec{y}^{\ast})$ that we were looking for. 
For ease of notation, once the values $( \vec{x}^{\,\ast}, \vec{y}^{\,\ast})$ are inserted into Eqs.~\eqref{eq:qij} and~\eqref{eq:pij}, we denote the resulting key probabilities as $q^\ast_{ij}$ and $p^\ast_{ij}$ respectively.

\subsection{The local filter\label{sec:flocal}
}
We can now introduce our filtering technique. We consider the local (and, as we argue later, most appropriate) version first, and then move on to the global one.
As we anticipated, our local filtering method is similar in spirit to the Disparity Filter (DF) introduced by Serrano \etal \cite{serrano-pnas-2009}, 
as it is based on the calculation of a $p$-value $\gamma^\ast_{ij}$ for each observed link of weight $w_{ij}^\ast>0$, defined as the probability that the null model produces a weight $w_{ij}\ge w_{ij}^\ast$, and on the removal of links for which $\gamma_{ij}^\ast$ is higher than a fixed critical value $\widetilde{\gamma}$.
However, our method improves upon the DF by recalculating the $p$-values according to the maximum-entropy probability $q_{ij}^*$ derived above, as we now describe.

Since in our case weights are discrete, $p$-values should be calculated by replacing the integral appearing in Eq.~\eqref{eq:serrano} with a sum:
\begin{eqnarray}
\gamma_{ij}^{\ast} &\equiv& \textrm{Prob}(w_{ij} \geq w_{ij}^{\ast}) \nonumber\\
&=& \sum_{w \geq w_{ij}^{\ast}} q^{\ast}_{ij} (w) \nonumber\\
&=& \left\{
    \begin{array}{ll}
      1 & \textrm{if  } w_{ij}^{\ast} = 0 \\
      1 - \sum_{w = 0}^{w_{ij}^{\ast}-1} q^{\ast}_{ij}(w) & \textrm{if  } w_{ij}^{\ast} > 0
     \end{array}
  \right. .
\end{eqnarray}
If a link is actually present in the observed network, \ie $w_{ij}^{\ast} > 0$, we have
\begin{eqnarray}
\gamma^{\ast}_{ij} & = & 1 - \sum_{w = 0}^{w_{ij}^{\ast}-1} q^{\ast}_{ij}(w)  \nonumber \\
            & = & 1 - \sum_{w = 0}^{w^{\ast}-1} \dfrac{\left( \xxi^{\ast} \xxj^{\ast} \right)^{\Theta(w)} \left( \yyi^{\ast} \yyj^{\ast} \right)^{w}  \left( 1 - \yyi^{\ast} \yyj^{\ast} \right) }{1 - \yyi^{\ast} \yyj^{\ast} + \xxi^{\ast} \xxj^{\ast} \yyi^{\ast} \yyj^{\ast}} \nonumber \\
            & = & 1 - \dfrac{ 1 - \yyi^{\ast} \yyj^{\ast} }{1 - \yyi^{\ast} \yyj^{\ast} + \xxi^{\ast} \xxj^{\ast} \yyi^{\ast} \yyj^{\ast}} \left[ 1 + \xxi^{\ast} \xxj^{\ast} \sum_{w = 1}^{w^{\ast}-1} \left( \yyi^{\ast} \yyj^{\ast} \right)^{w} \right]  \nonumber \\
            & = &  \dfrac{\xxi^{\ast} \xxj^{\ast} \left( \yyi^{\ast} \yyj^{\ast} \right)^{w_{ij}^{\ast}}}{1 - \yyi^{\ast} \yyj^{\ast} + \xxi^{\ast} \xxj^{\ast} \yyi^{\ast} \yyj^{\ast}}\nonumber\\
&=&p^\ast_{ij}\left( \yyi^{\ast} \yyj^{\ast} \right)^{w_{ij}^{\ast}-1}.
\label{eq:gamma_ecm}
\end{eqnarray}
The above quantity represents the probability of generating a link between nodes $i$ and $j$ with a weight equal to, or greater than, the observed weight $w_{ij}^\ast$. It can be seen from Eq.~\eqref{eq:gamma_ecm} that this probability coincides with the probability $p^\ast_{ij}$ that a link of unit weight is established, times the probability $\left( \yyi^{\ast} \yyj^{\ast} \right)^{w_{ij}^{\ast}-1}$ that the weight is successfully incremented $w_{ij}^{\ast}-1$ times (so that the total weight is at least $w_{ij}^{\ast}$), irrespective of whether possible attempts to further increment the weight beyond $w_{ij}^{\ast}$ are successful or not.

\emph{Per se}, $\gamma^\ast_{ij}$ represents the $p$-value associated with the null hypothesis that the edge weight $w_{ij}^\ast$ has been produced by mere chance, given the empirical strength and degree sequences $\vec{s}^\ast$ and $\vec{k}^\ast$. 
Links with a higher value of $\gamma^\ast_{ij}$ are closer to compatibility with the null hypothesis.
Therefore the quantity $1/\gamma^\ast_{ij}>0$ can be viewed as a rescaling of the original weight $w_{ij}^\ast>0$ that effectively reduces the absolute importance of large weights, if these are found between nodes with large strengths and/or degrees.
In principle, this rescaling can already be considered a form of filtering, that keeps all edges but with modified weights. 
In practice, we are going to fix a threshold value (corresponding to a desired level of statistical significance) and retain only the edges for which $1/\gamma^\ast_{ij}$, rather than $w^\ast_{ij}$, is larger than the threshold.
As we now show, this crucial step effectively replaces the problematic enforcement of a global threshold on the original weights with the enforcement of a local threshold that controls for the strengths and degrees of nodes.
In particular, we reject the null hypothesis, and therefore retain the observed link between nodes $i$ and $j$ as statistically significant, if $\gamma_{ij}^\ast$ is smaller than a desired threshold $\widetilde{\gamma}$:
\begin{equation}
\gamma^\ast_{ij}<\widetilde{\gamma}.
\label{eq:pvalue}
\end{equation}
Equivalently, using Eq.~\eqref{eq:gamma_ecm} the above homogeneous (global) threshold $\widetilde{\gamma}$ for $\gamma^\ast_{ij}$ translates into the following heterogeneous (local) threshold $\widetilde{w}_{ij}$ for $w^\ast_{ij}$:
\begin{equation}
w_{ij}^\ast>1+\frac{\ln (\widetilde{\gamma}/p_{ij}^\ast)}{\ln(y_i^\ast y_j^\ast)}\equiv \widetilde{w}_{ij}.
\label{eq:wtilde}
\end{equation}

It should be noted that the term on the r.h.s. depends on $x_i^\ast$, $x_j^\ast$, $y_i^\ast$, $y_j^\ast$. In turn, these four parameters depend \emph{on the entire empirical strength and degree sequences} $\vec{s}^*$ and $\vec{k}^*$ through Eqs.~\eqref{eq:s} and~\eqref{eq:k}, or equivalently~\eqref{eq:ML} and~\eqref{eq:LL}.
So, unlike the DF, where the statistical significance of the observed edge weight $w^\ast_{ij}$ is assessed against a null model that (upon double-checking from the point of view of both $i$ and $j$) depends only on the endpoint properties $s^\ast_{i}$, $k^\ast_{i}$, $s^\ast_{j}$, $k^\ast_{j}$, here the statistical test for $w^\ast_{ij}$ depends on the degrees and strengths \emph{of all nodes in the network}. 
This is a desirable property, following from the maximum-entropy nature of our model whereby the specified constraints \emph{collectively} determine the probability of each graph, and ultimately each edge, in the ensemble.\\

Summing up, our local filtering method is very simple: given the empirical network ${\bm W}^\ast$ with strength sequence $\vec{s}^\ast$ and degree sequence $\vec{k}^\ast$, we 
\begin{itemize}
\item find the values $(\vec{x}^\ast,\vec{y}^\ast)$ through Eqs.~\eqref{eq:s} and~\eqref{eq:k}, or equivalently~\eqref{eq:ML} and~\eqref{eq:LL} (efficient algorithms serving this purpose have been devised~\cite{squartini-njp-2015} and coded~\cite{mathworks,eli}); 
\item retain only the links (along with their weight $w^\ast_{ij}$) that realize Eq.~\eqref{eq:pvalue}, or equivalently~\eqref{eq:wtilde}, for a given value of the threshold $\widetilde{\gamma}$
(a generally accepted reference choice is $\widetilde{\gamma}=0.05$, although we will show results for a wide range of values of $\widetilde{\gamma}$).
\end{itemize}
We refer to the resulting pruned network as the \emph{local backbone} of ${\bm W}^\ast$ and denote it in terms of the ($\widetilde{\gamma}$-dependent) matrix
${\bm \varSigma}^\textrm{local}(\widetilde{\gamma})$ with entries $\sigma_{ij}^\textrm{local}(\widetilde{\gamma})$.
Clearly, the extreme cases are ${\bm \varSigma}^\textrm{local}(1)={\bm W}^\ast$ (all links of the original network being preserved) and ${\bm \varSigma}^\textrm{local}(0)={\bm 0}$, the latter denoting a matrix will all zero entries, \ie an empty graph.

\subsection{The global filter\label{sec:fglobal}}

So far, we have implemented the filter locally by computing the significance of each link $\gamma_{ij}$ and comparing it with a given critical $p$-value $\widetilde{\gamma}$. However, in analogy with~\cite{dianati-pre-2016}, it is in principle possible to apply the same filter in a global, whole-graph fashion. In this context, we assume that the most significant \emph{global backbone} $\bm{\varSigma}^\textrm{global}(\widetilde{L})$ with $ \widetilde{L}\le L^\ast$ links (where $L^\ast$ is the empirical number of links in the original network $\bm{W^{\ast}}$) is the minimum-likelihood subgraph among the set of all subgraphs of the original network having $\widetilde{L}$ edges. 
This is equivalent to claiming that $\bm{\varSigma}^\textrm{global}(\widetilde{L})$ is the subnetwork which is least likely to be generated by pure chance. For each weighted subgraph $\bm{\varSigma}$ of the observed network $\bm{W^{\ast}}$, the likelihood is
\begin{eqnarray}
\label{eq:prob_glob}
P(\bm{\varSigma}|\bm{W^{\ast}}) = \prod_{i < j}^{} \left[ q_{ij} (\sigma_{ij})  \right]^{a^\ast_{ij}} = \prod_{i < j}\left[ q_{ij} (w_{ij}^\ast)  \right]^{a^\ast_{ij}} \,,
\end{eqnarray}
where $\sigma_{ij}$ is the weight of the link between $i$ and $j$ in the subgraph $\bm{\varSigma}$ and $a^\ast_{ij} = 0,1$ is the element of the adjacency matrix of the original graph $\bm{W^{\ast}}$. 
The global backbone (for given $\widetilde{L}$) is then defined as
\begin{equation}
\bm{\varSigma}^\textrm{global}(\widetilde{L}) = \underset{\bm{\varSigma}:L(\bm{\varSigma})=\widetilde{L}}{\mathrm{argmin}}\,P(\bm{\varSigma}|\bm{W^{\ast}}),
\end{equation}
where $L(\bm{\varSigma})$ denotes the number of links in the subgraph $\bm{\varSigma}$.

Given $\widetilde{L}$, the minimum of the likelihood is achieved by the $\widetilde{L}$ smallest factors of the product in Eq.~\eqref{eq:prob_glob}. 
Hence, the entries of $\bm{\varSigma^{\ast}}(\widetilde{L})$ are easily found to be
\begin{eqnarray}
\sigma_{ij}^\textrm{global}(\widetilde{L}) = \left\{
     \begin{array}{cl}
        w^\ast_{ij} & \mbox{if } (i,j) \in \lambda_{\widetilde{L}} (\bm{W^{\ast}}) \\
       0      & \mbox{otherwise}
     \end{array} 
    \right.,\nonumber
\end{eqnarray}
where $ \lambda_{\widetilde{L}} (\bm{W^{\ast}}) $ is the set of the $\widetilde{L}$ least likely links, \ie those with the smallest probabilities $ q_{ij} (w^\ast_{ij}) $. 

Note that, while the local filter selects links based on statistical significance, this is not the case for the global one. Nonetheless, it is worth comparing the local backbone, for a given $\widetilde{\gamma}$, with the global one obtained using a value of $\widetilde{L}$ giving as many links as the local backbone.
This effectively establishes a relationship between $\widetilde{L}$ and $\widetilde{\gamma}$.
It then becomes clear that the difference between the local filter and the global one is the fact that the latter selects the $\widetilde{L}$ links for which the probability mass function $\textrm{Prob}(w_{ij} =w_{ij}^{\ast})$ is minimum, while the former selects the $\widetilde{L}$ links for which the \emph{cumulative} probability function $\textrm{Prob}(w_{ij} \ge w^\ast_{ij})$ is minimum. 
One can at this point note that, as in the usual construction of one-sided tests and the associated $p$-values in statistics, the use of the cumulative probability is much more reasonable, as it makes more sense to define compatibility with the null model in terms of the chance that the edge weight is equal to \emph{or larger than}, rather than only equal to, the empirical one.
We therefore claim that the local method should be preferred over the global one.
We also note that, if the global filter were redefined in terms of cumulative probability, the two methods would coincide. Nevertheless, in the following we measure also the performances of the two methods and present an empirical \emph{a posteriori} confirmation of our claim.

\section{Empirical analysis\label{sec:results}}

In this Section we gauge the performances of the ECM filter and compare its filtering power with that of the disparity and \gloss methods. All three methods are based on null models that, by preserving (among other properties) the degree sequence of the original network, automatically preserve the link density and therefore allow for a consistent comparison. 
We do not include the method by Dianati in this comparison since, as we mentioned, it does not preserve the link density and therefore tends to retain too many spurious links, many of which would be reducible to the knowledge of node degrees. After that, we compare the networks filtered with the local and global versions of ECM. Finally, we show how our filter is able to dig out interesting hidden patterns by presenting the results obtained for the time-varying World Trade and US airport networks.

\subsection{Data\label{sec:data}}
Here we provide a short description of the datasets used, and in Tab.~\ref{tab:datasets} we list their fundamental topological features.
\begin{basedescript}{%
  \desclabelstyle{\multilinelabel}
  \desclabelwidth{2cm}}
\setlength\itemsep{0.2cm}
 \item[Domestic flights in the U.S.A.] A node corresponds to an airport of U.S.A. and a link between two airports exists if there is a direct flight connecting them. The weight of a link indicates the number of passengers transiting between two airports \cite{barrat-pnas-2004}.
 \item[Florida Bay Foodweb] The network describes the trophic interactions between species during the dry season in the South Florida Bay ecosystem. The data have been collected from the ATLSS Project by the University of Maryland \cite{ulanowicz-food-1998}. Nodes correspond to species and links represent the carbon flows (mg C y$^{-1}$ m$^{-2}$) among them. 
 \item[Star Wars movies] The data portraits the interactions between the characters of the Star Wars films saga. Each node represents a character of the cast, while a link connects two characters if they both speak in the same scene and the weight counts the number of different scenes that they share across the whole seven episodes saga \cite{gabasova-starwars-2015}.
 \item[World Trade snapshots] The networks represent the trading volumes between countries in the period between the years 1998 and 2011. Each year is encoded as a distinct network. A node indicates a country, while the weight of a link denotes the gross trade volume (measured in thousands of US dollars) between two countries \cite{Comtrade, Baci}.
 \item[World Trade Multiplex] Multiplex representation of trade volumes for year 2011. Each layer represents a different commodity \cite{boccaletti-phys_rep-2014}, namely: Fish, crustaceans and acquatic invertebrates (FISH); Cereals (CER); Mineral fuels, mineral oils and products of their distillation, bitumin substances, mineral wax (FUEL/OIL); Iron and steel (IRON) \cite{Comtrade, Baci}.
\end{basedescript}
\begin{table}[t]
\centering
\begin{tabular}[c]{l|ccc}
\hline\hline
\bf{Network} & $N$ & $L$ & $\rho \, (\%)$ \\\hline\hline
\emph{US aiports network} & 426 & 2439 & 2.69 \\
\hline
\emph{Florida Bay foodweb} & 126 & 1969 & 25.00 \\
\hline
\emph{Star Wars network: }&&&\\
All merged & 111 & 444 & 7.27 \\
All & 112 & 450 & 7.24 \\
Full int & 110 & 398 & 6.64 \\
Mentions & 113 & 817 & 12.91 \\\hline
\emph{World trade network: }&&& \\
Year 1998 & 208 & 10210 & 47.43 \\
Year  1999 & 208 & 10904 & 50.65 \\
Year  2000 & 208 & 11778 & 54.71 \\
Year  2001 & 208 & 12256 & 56.93 \\
Year  2002 & 208 & 12523 & 58.17 \\
Year  2003 & 208 & 12796 & 59.44 \\
Year  2004 & 208 & 12921 & 60.02 \\
Year  2005 & 208 & 13145 & 61.06 \\
Year  2006 & 208 & 13146 & 61.06 \\
Year  2007 & 208 & 13230 & 61.45 \\
Year  2008 & 208 & 13489 & 62.66 \\
Year  2009 & 208 & 13360 & 62.06 \\
Year  2010 & 208 & 13321 & 61.88 \\
Year  2011 & 208 & 12956 & 60.18 \\\hline
\emph{World trade multiplex:  }&&& \\
 Fish & 207 & 4628 & 21.71 \\
 Cereals & 207 & 3474 & 16.29 \\
 Fuel/oil & 207 & 5711 & 26.79 \\
 Iron & 207 & 5348 & 25.08 \\\hline\hline
\end{tabular}
\caption{Topological characteristics of the datasets used. For each network, we report the number of nodes $N$, of edges $L$ and the link density $\rho$ for the non filtered case.}
\label{tab:datasets}
\end{table}
\begin{figure*}[t]
\centering
\includegraphics[width=1.0\linewidth]{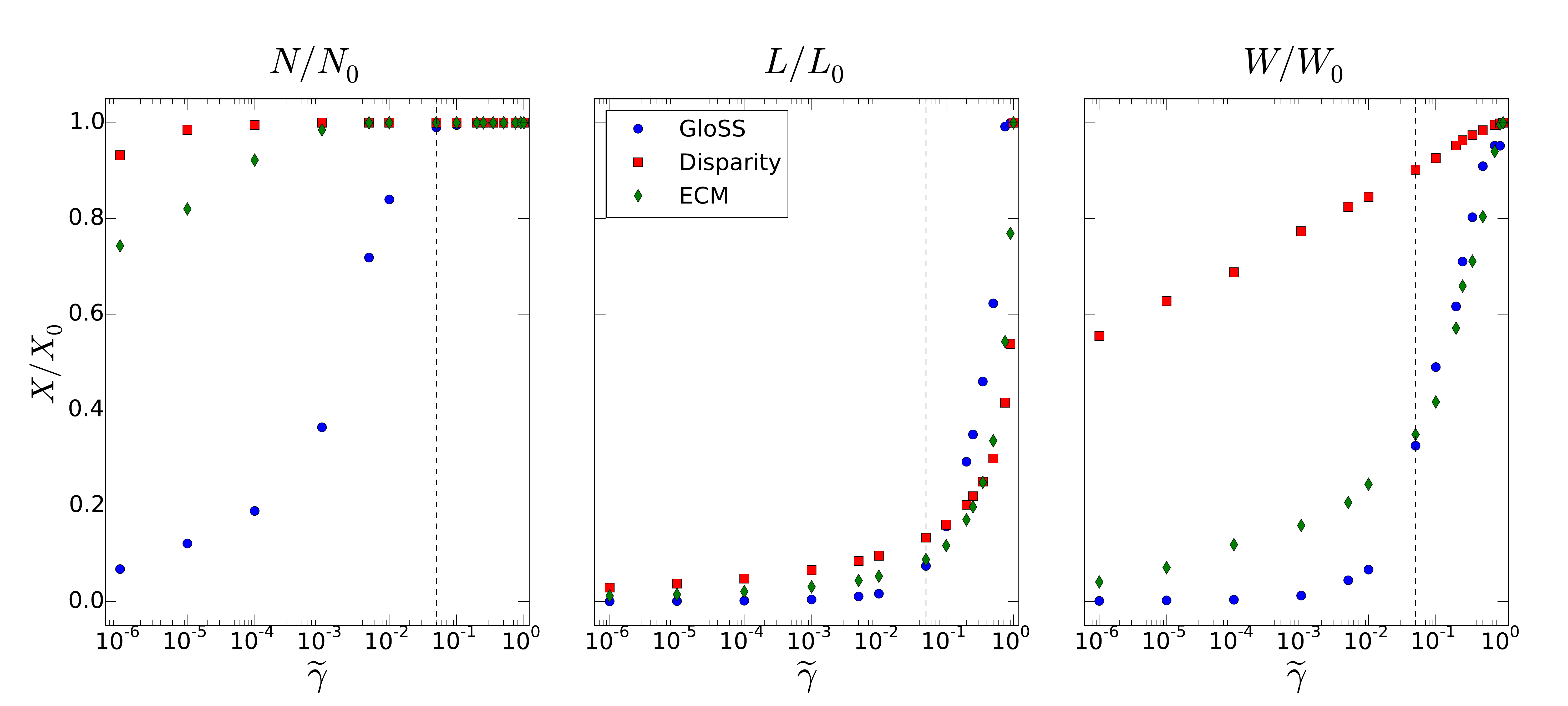}
\caption{Effects of filtering on three topological indicators, for the International Trade Network in 2011: fraction of nodes ${N}/{N_0}$ (left), links ${L}/{L_0}$ (center), and total weight ${W}/{W_0}$ (right) as a function of the $p$-value $\widetilde{\gamma}$. Red dots refer to the disparity filter, blue to GloSS, and green to ECM.} 
\label{fig:methods-compar}
\end{figure*}
\subsection{Typical results and comparison with other methods}
When it comes to performances, a good filtering technique should, ideally, be able to prune as many connections as possible while preserving the highest amount of information and avoiding the breakup of the system. A good way to measure such ability is computing the fraction of a given quantity $X$ preserved after filtering, ${X}/{X_0}$, where $X_0$ denotes the same quantity measured in the original network. We consider three indicators: number of nodes ($N$), of edges ($L$) and the total weight ($W$) respectively. The behaviour of these indicators for different filtering intensities (\ie the $p$-values) for the International Trade Network in 2011 is displayed in Fig.~\ref{fig:methods-compar}.

In the left panel, we notice how all methods return networks with no isolated nodes up to $\widetilde{\gamma} \simeq 0.06$. Below such value, the disparity filter appears to be the most conservative method because its local nature tends to avoid the pruning of all the connections of a node. 
At the other extreme, despite imposing the conservation of the initial topology, GloSS is the most aggressive method, isolating more than 20\% of nodes for $\widetilde{\gamma} < 0.05$. 
The ECM filter, instead, stays in between these boundaries and achieves a trade-off between its aggressive and conservative counterparts. In the strong filtering regime, which corresponds to the typical accepted range of $p$-values $\widetilde{\gamma} < 0.05$, the established hierarchy holds also for ${L}/{L_0}$ and ${W}/{W_0}$. The scenario changes, instead, for $\widetilde{\gamma} > 0.05$. In this regime ECM prunes out more connections than \gloss (central panel) as well as heavier than disparity ones (right panel). More specifically, since \gloss redistributes only the weights keeping the topology unaltered, this artificially boosts the significance of each link making it harder to remove. The stark difference in the behaviour of ${W}/{W_0}$ for the DF is, instead, the hallmark of bias towards heavier edges. Although the preservation of heavy connections might seem an advantage this is not always the case. As we will show later, heavier connections tend to conceal interesting features of the system such as its mesoscopic structure (like, for instance, the presence of communities \cite{fortunato-phys_rep-2010, liebig-epl-2016}).

\subsection{Local versus global filtering}
\begin{figure*}[t]
\centering
\includegraphics[width=0.8\linewidth]{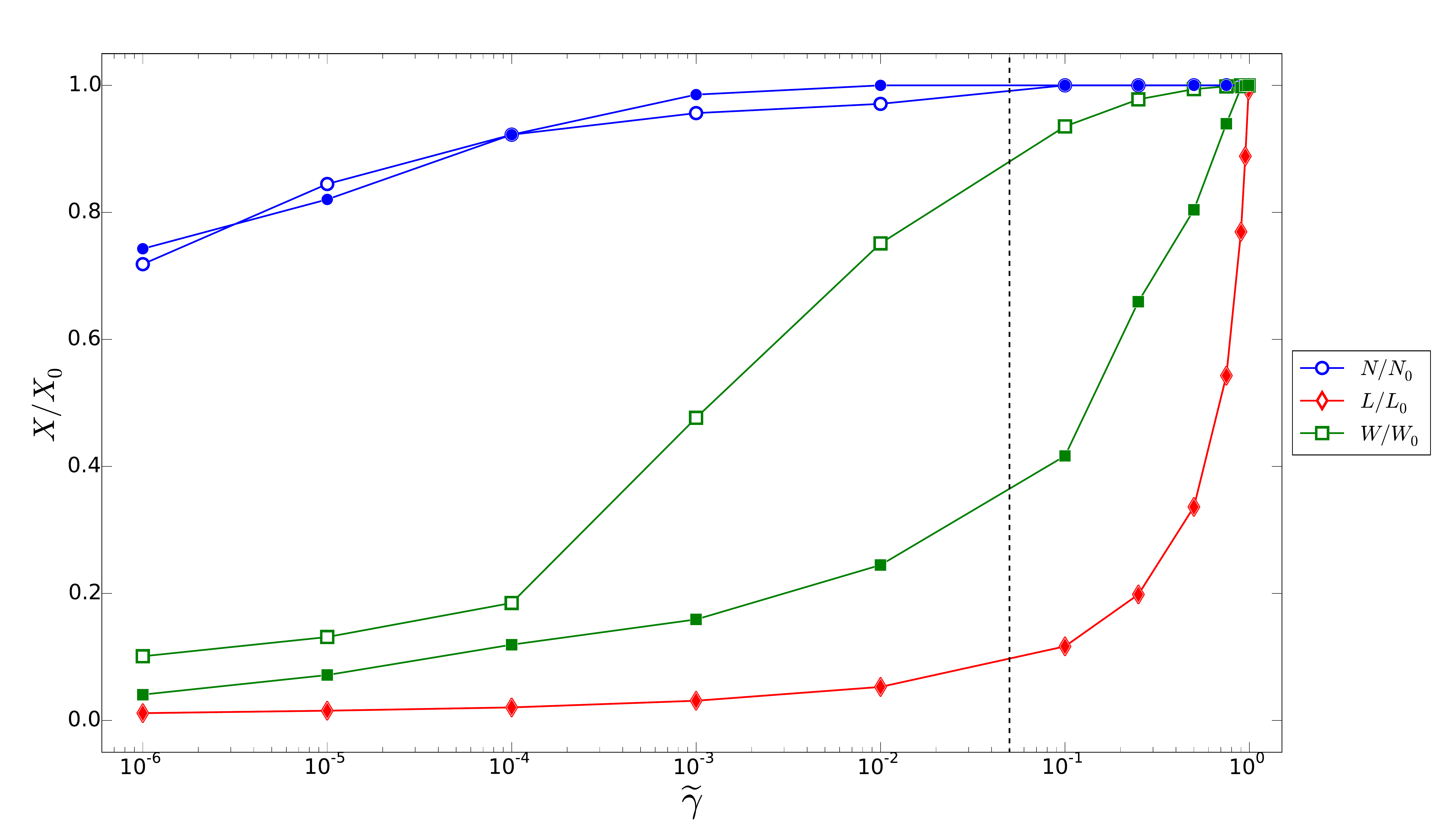}
\caption{Effects of the global and local implementations of the ECM filter on three topological indicators: fraction of nodes ${N}/{N_0}$ (blue dots), links ${L}/{L_0}$ (red diamonds), and total weight ${W}/{W_0}$ (green squares) as a function of the $p$-value $\widetilde{\gamma}$. Filled symbols refer to the local filter and empty symbols to the global one.}
\label{fig:compare_local_global}
\end{figure*}

Besides comparing the ECM filter with other existing solutions, it is worth comparing also its \emph{global} and \emph{local} versions. Given a certain $p$-value $\widetilde{\gamma}$, to properly compare the two implementations we first produce the local backbone, which results in a certain number $\widetilde{L}$ of links, and then rank all the edges of the original network according to their link probability $q^*_{ij}$ using Eq.~\eqref{eq:qij}.
We thus obtain the global backbone by retaining only the first $\widetilde{L}$ links. 
Finally, to study the differences between the two filtering approaches, in Fig.~\ref{fig:compare_local_global} we display the fraction of nodes, edges and total weight with respect to the $p$-value for the International Trade Network in 2011. By construction, the trends showing the fractions of retained links coincide. Furthermore, the analysis of Fig.~\ref{fig:compare_local_global} denotes no qualitative difference between the fraction of preserved nodes in the two methods. This is however not the case when we consider the residual total weight: indeed, we observe that the global ECM filter preserves significantly more weight than the local one, in particular for $p$-values higher than $ 10^{-4}$.

\begin{figure*}[t]
\centering
\includegraphics[width=1.\linewidth]{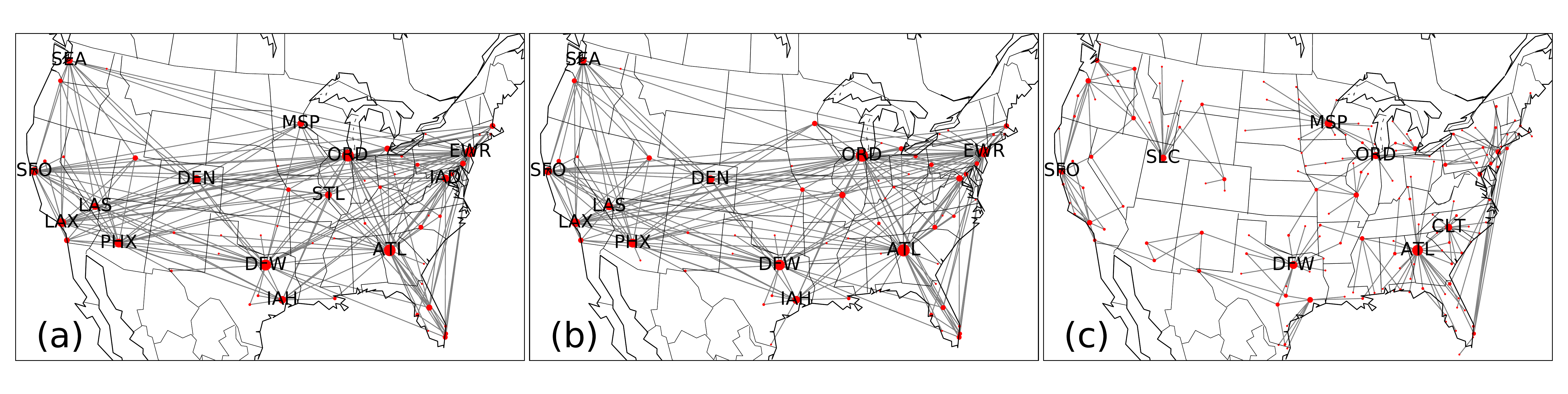}
\caption{Visual representation of the 200 heaviest links in the original network (left), ECM global filter (center) and ECM local (right) for the US Airport Network. The local filtering is obtained using $\widetilde{\gamma} = 0.05$ and the global filtering is constructed such that the number of links is the same as in the local one.}
\label{fig:maps_local_global_heaviest_usair}
\end{figure*}

A portrait of the differences between the local and global filter can be found in Fig.~\ref{fig:maps_local_global_heaviest_usair} and Tab.~\ref{tab:links_glob_loc_wei_usair} (results for other datasets can be found in the Supplementary Materials), where we display the case of US airports dataset. The most striking feature of Fig.~\ref{fig:maps_local_global_heaviest_usair} is the stark difference between the local network (right panel) and the other two. In the local network, in fact, we observe the emergence of a clear hub-and-spoke pattern \cite{bryan-jregsci-1999}. Indeed, the list of the 20 heaviest edges (Tab.~\ref{tab:links_glob_loc_wei_usair}) confirms that there is not very much difference between the global network and the original one in terms of backbone, while the difference becomes much stronger in the local case with the appearance of many connections among global ``tier-1'' hubs like New York and San Francisco and ``tier-2'' airports like Austin, Cleveland and Indianapolis, just to name a few as clearly shown in the graphs displayed in Fig.~\ref{fig:maps_local_global_heaviest_usair}.

\begin{table*}[t]
\begin{center}
 \begin{tabular}{||c | c | c | c||} 
 \hline
 Rank & Orginal network & Local filter & Global filter \\ [0.5ex] 
 \hline\hline
 1 & Los Angeles - San Francisco & Las Vegas - Los Angeles & Los Angeles - San Francisco \\ 
 \hline
 2 & Las Vegas - Los Angeles & Boston - New York & Las Vegas - Los Angeles \\
 \hline
 3 & Los Angeles - Phoenix & Seattle - San Francisco & Los Angeles - Phoenix \\
 \hline
 4 & New York - Chicago & New York - Ft Lauderdale & New York - Chicago \\
 \hline
 5 & Los Angeles - Chicago & San Diego - San Francisco & Los Angeles - Chicago \\  
 \hline
 6 & Dallas - Houston & Los Angeles - Sacramento & Dallas - Houston \\ 
 \hline
 7 & New York - Los Angeles & Portland - San Francisco & New York - Los Angeles \\
 \hline
 8 & Chicago - San Francisco & Houston - New Orleans & Chicago - San Francisco \\
 \hline
 9 & Atlanta - New York & Kansas City - Chicago & Atlanta - New York \\
 \hline
 10 & Boston - New York & Dallas - San Antonio & Boston - New York \\ 
 \hline
 11 & New York - Washington & Austin - Dallas & New York - Washington \\
 \hline
 12 & Dallas - Los Angeles & Houston - San Antonio & Dallas - Los Angeles \\
 \hline
 13 & Seattle - San Francisco & Austin - Houston & Seattle - San Francisco \\
 \hline
 14 & Las Vegas - San Francisco & Cleveland - Chicago & Las Vegas - San Francisco \\ 
 \hline
 15 & New York - San Francisco & New York - West Palm Beach & New York - San Francisco \\
 \hline
 16 & New York - Ft Lauderdale & Albuquerque - Phoenix & New York - Ft Lauderdale \\
 \hline
 17 & Minneapolis St Paul - Chicago & Spokane - Seattle & Minneapolis St Paul - Chicago \\ 
 \hline
 18 & San Diego - San Francisco & Indianapolis - Chicago & San Diego - San Francisco \\
 \hline
 19 & Los Angeles - Sacramento & Atlanta - Jacksonville & Los Angeles - Sacramento \\
 \hline
 20 & Denver - Chicago & Reno - San Francisco & Denver - Chicago \\ 
 \hline
\end{tabular}
\end{center}
\caption{List of the 20 heaviest links in the US Airport Network in the original network (left column), and after applying the local (center) and global (right) ECM filters.}
\label{tab:links_glob_loc_wei_usair}
\end{table*}

\begin{figure*}[t]
%
\includegraphics[width=0.75\linewidth]{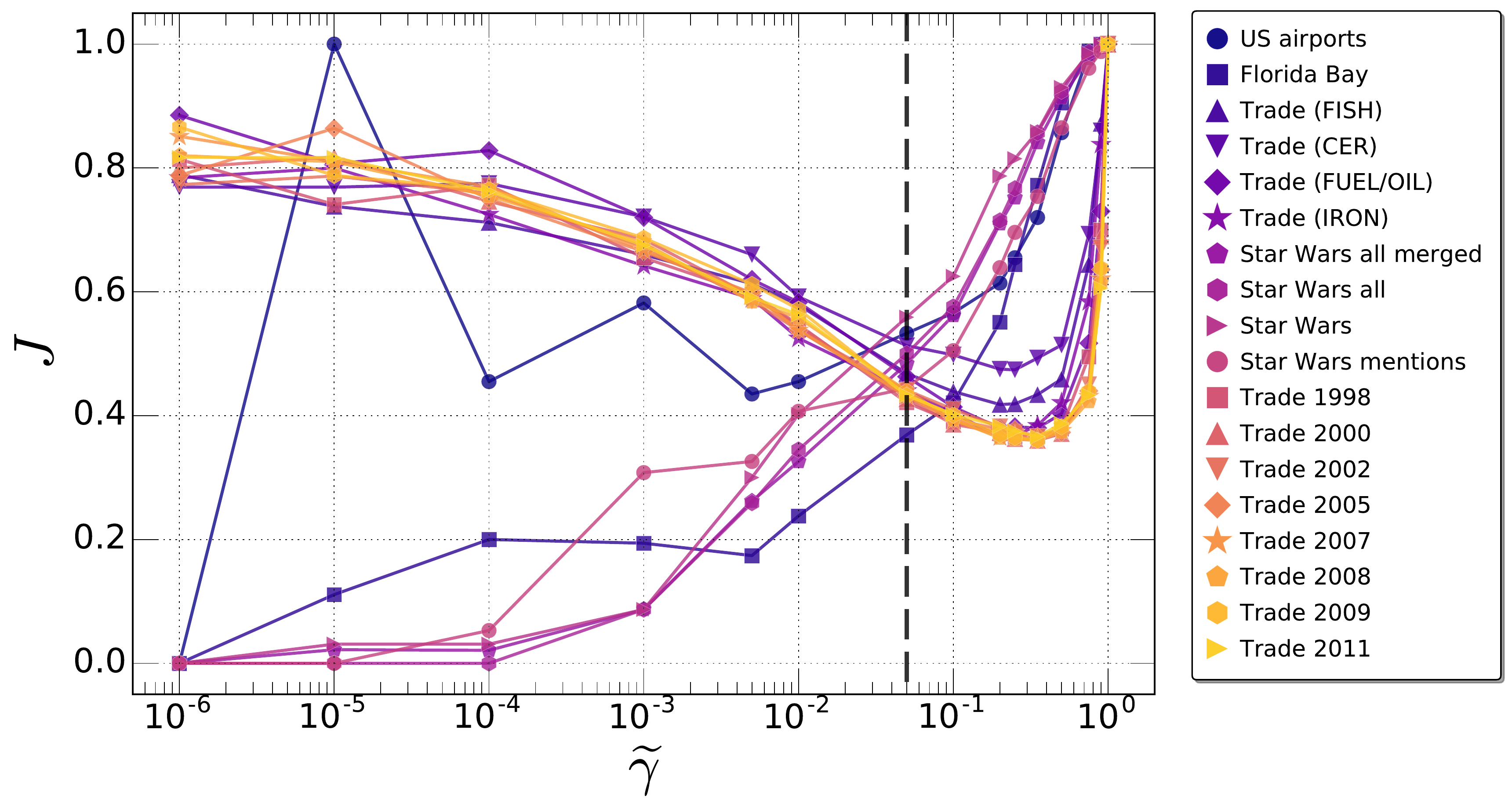}
\caption{Jaccard score $J$ of local and global filtered backbones obtained at different $p$-values $\widetilde{\gamma}$ for all the datasets considered in our study. The vertical dashed line denotes $\widetilde{\gamma} = 0.05$.}
\label{fig:jaccard_all}
\end{figure*}

A more detailed analysis of the similarity between the local and global networks as a function of the $p$-value is provided by computing the Jaccard score $J$ \cite{jaccard_score-1902}. This score quantifies the similarity between two sets $A$ and $B$ by computing the ratio between the cardinality of the intersection and the cardinality of the union, \ie
\begin{equation}
\label{eq:jaccard}
J = \dfrac{\lvert A \cap B \rvert}{\lvert A \cup B \rvert} \,.
\end{equation}
A value $J=1$ indicates that $A$ and $B$ are exactly the same set, while a value $J=0$ denotes that the sets are completely different. In our case, we calculate $J$ for the sets of edges belonging to the local and global networks computed using different values of $\widetilde{\gamma}$. The results for all the datasets are visible in Fig.~\ref{fig:jaccard_all}. The shape of $J$ versus $\widetilde{\gamma}$ highlights two distinct behaviours. 
In one case, the similarity between local and global networks tends to fade away monotonically as we increase the aggressivity of the filtering. 
In the other case, the two networks initially tend to differentiate and become more and more alike thereafter. 
The World Trade Network is an example of the latter behaviour. As we increase the aggressivity of the filter, we observe the presence of a minimum of similarity around $\widetilde{\gamma} \simeq 0.35$ followed by an increase up to $J \approx 0.8$ for $\widetilde{\gamma} = 10^{-6}$. One culprit of such behaviour is that the original networks are, in general, denser than the others ($\langle \rho \rangle \gtrsim 58\%$ for time-varying and $\langle \rho \rangle \gtrsim 22\%$ for single commodities). For $\widetilde{\gamma} > 0.35$ (which corresponds to a low level of statistical significance), the ECM filters (local and global) tend to prune out different links as suggested by the decrease of $J$. Instead, the behavior of $J$ for $\widetilde{\gamma} \leq 0.35$ (which includes all acceptable ranges of significance) suggests the emergence of a backbone shared by both networks which is resilient to pruning, resulting in an increase of $J$.


\subsection{The filter at work on multiplex networks}
The presence of a similar trend in the Jaccard score between global and local backbones of yearly and single commodities leads us to investigate the effect that the aggregation of multiple commodities has on the extraction of the backbone. A \emph{multiplex} network representation \cite{de_domenico-prx-2013, boccaletti-phys_rep-2014} provides the natural way to study such effect. It has been proven, in fact, that the topological properties of single layers and aggregate networks may differ a lot \cite{cardillo-scirep-2013}; whilst in some cases, the multiplex  can be \emph{reduced}, deleting entire layers without losing information \cite{de_domenico-natcomm-2015}. It is therefore reasonable to ask whether filtering the layers first, and projecting them onto a single layer then, produces a filtered backbone whose structural properties are different from those of the network obtained inverting the order of these operations. 

\begin{figure*}[t]
\centering
\includegraphics[width=0.9\linewidth]{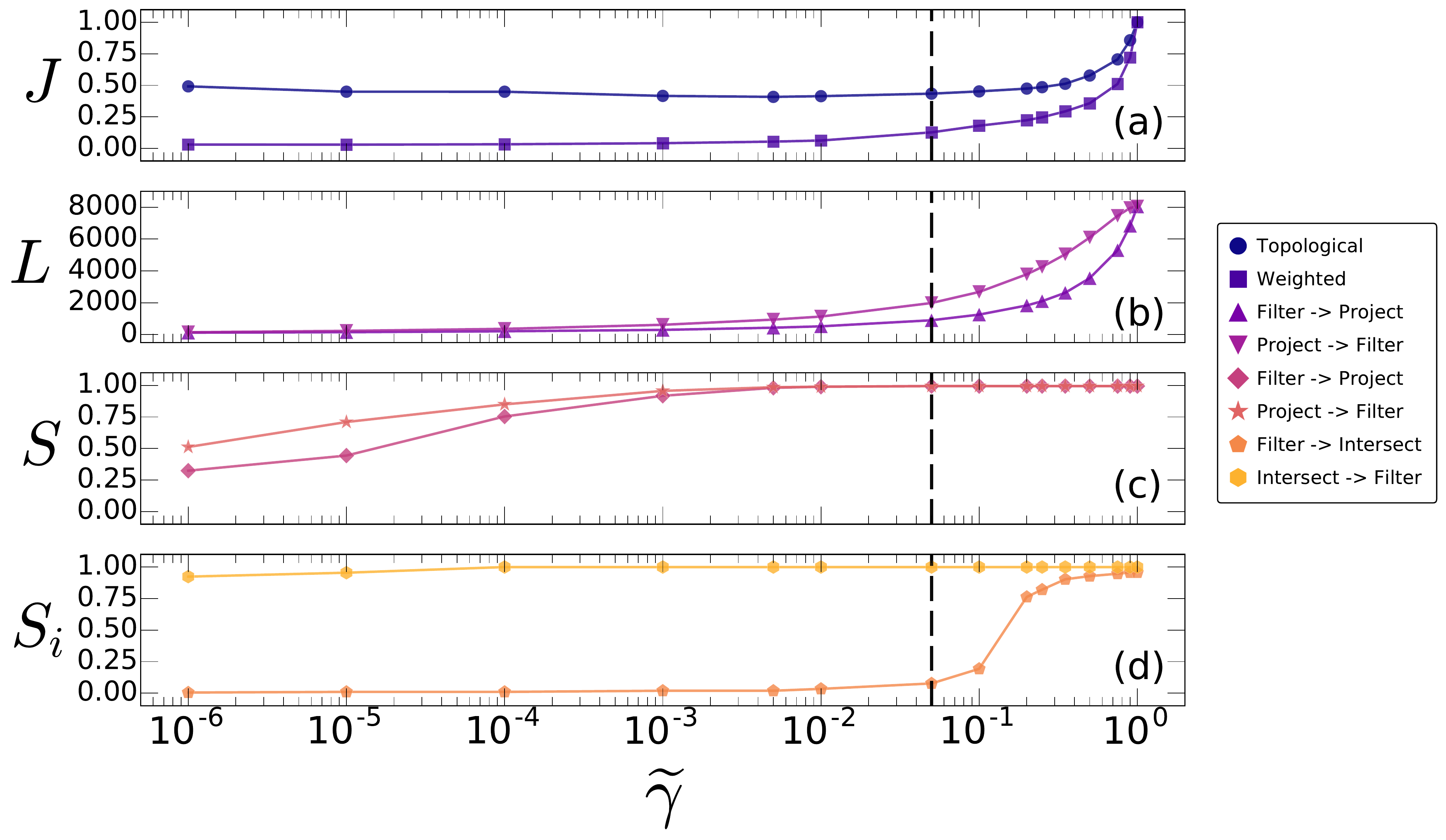}
\caption{Effect of multiplexity on the extraction of the weighted backbone. From top to bottom, we report the Jaccard score $J$ (a), number of edges $L$ (b), size of the giant component $S$ (c) and size of the mutually connected component $S_i$ as a function of $p$-value $\widetilde{\gamma}$. The quantities $L$, $S$ and $S_i$ are computed in networks obtained filtering each layer first and projecting them then (Filter $\rightarrow$ Project/Intersect) or in the inverse order case (Project/Intersect $\rightarrow$ Filtering).}
\label{fig:multiplex_quantities}
\end{figure*}

In Fig.~\ref{fig:multiplex_quantities}, we report the evolution of four topological indicators, namely: Jaccard score ($J$), number of edges ($L$), size of the giant component ($S$) and size of the mutually connected component ($S_i$), with respect to $\widetilde{\gamma}$. In Fig.~\ref{fig:multiplex_quantities}a we display the similarity of the backbones using the Jaccard score $J$. We can gauge the similarity either in a topological sense (the same link existing in both backbones) or in a weighted one (the link existing in both backbones with the same weight $w$). Except for the case where no filtering is performed, the weighted similarity is always smaller than the topological one, suggesting that the connection among the same countries is significative for one specific commodity but not in the remaining ones. Additionally, after an initial increase, for $\widetilde{\gamma} < 0.01$ the difference between the topological and weighted similarities remains more or less constant. In general, we observe that filtering before projecting returns a network which has fewer edges $L$ (Fig.~\ref{fig:multiplex_quantities}b) and a smaller giant component $S$ (Fig.~\ref{fig:multiplex_quantities}c).\\

The existence of a giant component is crucial for the appearance of several collective phenomena like synchronization and spreading, just to cite a few \cite{barrat-book-2008}. 
In multiplex networks, besides the giant component in the single layers and in the aggregate network, the so-called \emph{mutually connected component} (here defined as the network obtained projecting only the edges appearing in all the layers) plays a key role in the emergence of collective phenomena as well~\cite{buldyrev-nature-2010}. 
In Fig.~\ref{fig:multiplex_quantities}d, we compute the size of the mutually connected component, $S_i$, of the networks obtained filtering the layers first, extracting the intersection of the edge sets then and projecting them finally (orange pentagons).
We also show the result obtained by computing the intersection first, projecting the layers and filtering the aggregate network then (yellow hexagons). As we can see from panel d, the behavior of these two quantities with respect to $\widetilde{\gamma}$ is completely different. In particular, for $\widetilde{\gamma} < 0.2$ one case is above the critical percolation threshold while the other is already completely fragmented \cite{stauffer-book-1994}. Moreover, a visual representation of the original networks and their respective backbones for some commodities is available in the Supplementary Materials in Figs.~\ref{fig:maps_local_BACI2011_comm03_comm10} and \ref{fig:maps_local_BACI2011_comm27_comm72}.
%


\subsection{`Irreducible patterns' revealed by the method}
Finally, we illustrate several results showing that the ECM-filtered backbones can unveil significant information  about real-world systems that would otherwise remain hidden or not completely revealed by the other methods. 
We provide also some interpretation of the uncovered patterns. 
For the sake of brevity, here we discuss only the results obtained for the US airports and the International Trade networks, albeit similar conclusions can be drawn from the analysis of the other datasets as well, as shown in the Supplementary Materials. 

\begin{figure*}[t]
\centering
\includegraphics[width=\textwidth]{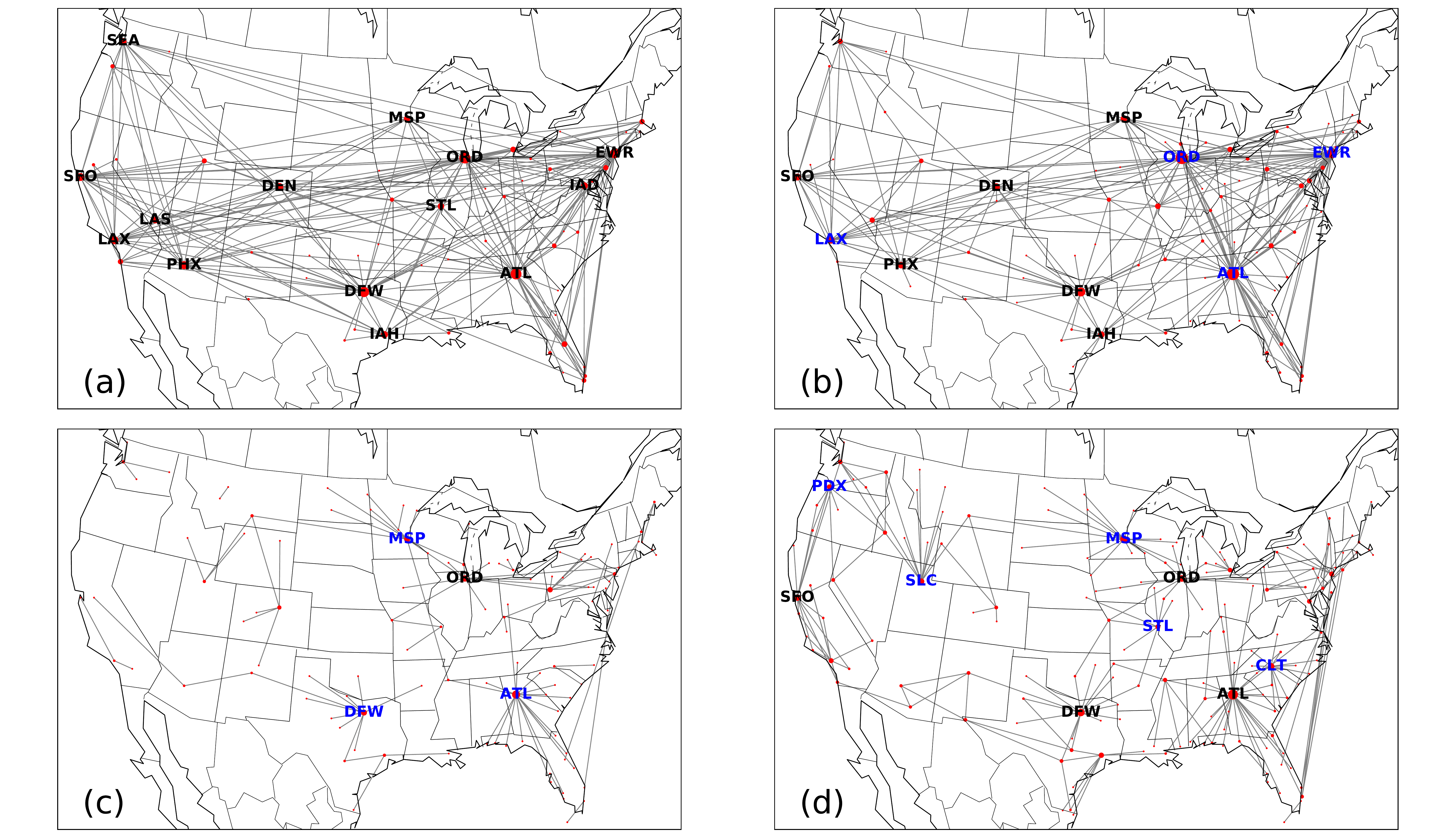}
\caption{Filtered backbones of the US airport network. In each map, we display the top 200 heaviest connections for the original (a), disparity (b), \gloss (c) and ECM (d) networks. All the filtered backbones have been obtained using a common $p$-value equal to $\widetilde{\gamma}=0.05$.}
\label{fig:airports-methods}
\end{figure*}

We start from US airports and refer to Fig.~\ref{fig:airports-methods}, where we show the original network (panel a) compared with the results of three filtering techniques: disparity (panel b), \gloss (panel c), and the local ECM filter (panel d). 
For all three methods, the backbones are extracted using always the same $p$-value $\widetilde{\gamma} = 0.05$ for consistency and, to facilitate visual comparison, we display only the first 200 heaviest connections. 
At first glance, we notice a stark difference between the three backbones. More specifically, the disparity backbone is akin to the original network, displaying several long-range connections between airports like Atlanta (ATL), Chicago (ORD), Newark (EWR) and Los Angeles (LAX) to mention a few. 
All these airports are among the top 12 in terms of the number of passengers in 2002 as reported by the Federal Aviation Administration (FAA) of the United States \cite{faa-data-2002}.
The pattern of connections resembles therefore a \emph{point-to-point} one \cite{barthelemy-physrep-2011}. Qualitatively, this is in agreement with the tendency of the disparity filter to preserve heavier connections as reported in Fig.~\ref{fig:methods-compar}.
This is not the case for \gloss (panel c) and ECM (panel d). The former returns a very sparse backbone having less than 200 edges (so all the retrieved edges are shown in this case) where, despite the emergence of some star-like structures centered around Atlanta (ATL), Minneapolis (MSP) and Dallas (DFW), there are almost no connections at all in the west of the country.
Such a result is clearly undesirable, at it would imply no relevant connections for many US states at the chosen $p$-value, resulting in a heavily fragmented network.
By contrast, the ECM filter displays a much more spread-out pattern consisting of several local hubs, not directly connected to each other. This corresponds to the so called \emph{hub-and-spoke}, a well known structure observed in many spatial systems and indeed used in the airline system \cite{bryan-jregsci-1999, barthelemy-physrep-2011}. 
The hub-and-spoke structure is usually the result of a design aimed at minimizing the operational cost, here emphasizing the role of regional hubs as Salt Lake City (SLC), Minneapolis (MSP), Portland (PDX), Charlotte (CLT) and St. Louis (STL), to name a few. 
In other words, the ECM filter uncovers the cost-oriented hub-and-spoke structure of US airports that is hidden within large-flow point-to-point patterns.
Importantly, all US states are connected in the ECM backbone, making the resulting structure overall connected and hence much more acceptable in terms of transportation constraints.
%
%
%
%

\begin{figure*}[t]
\centering
\includegraphics[width=\textwidth]{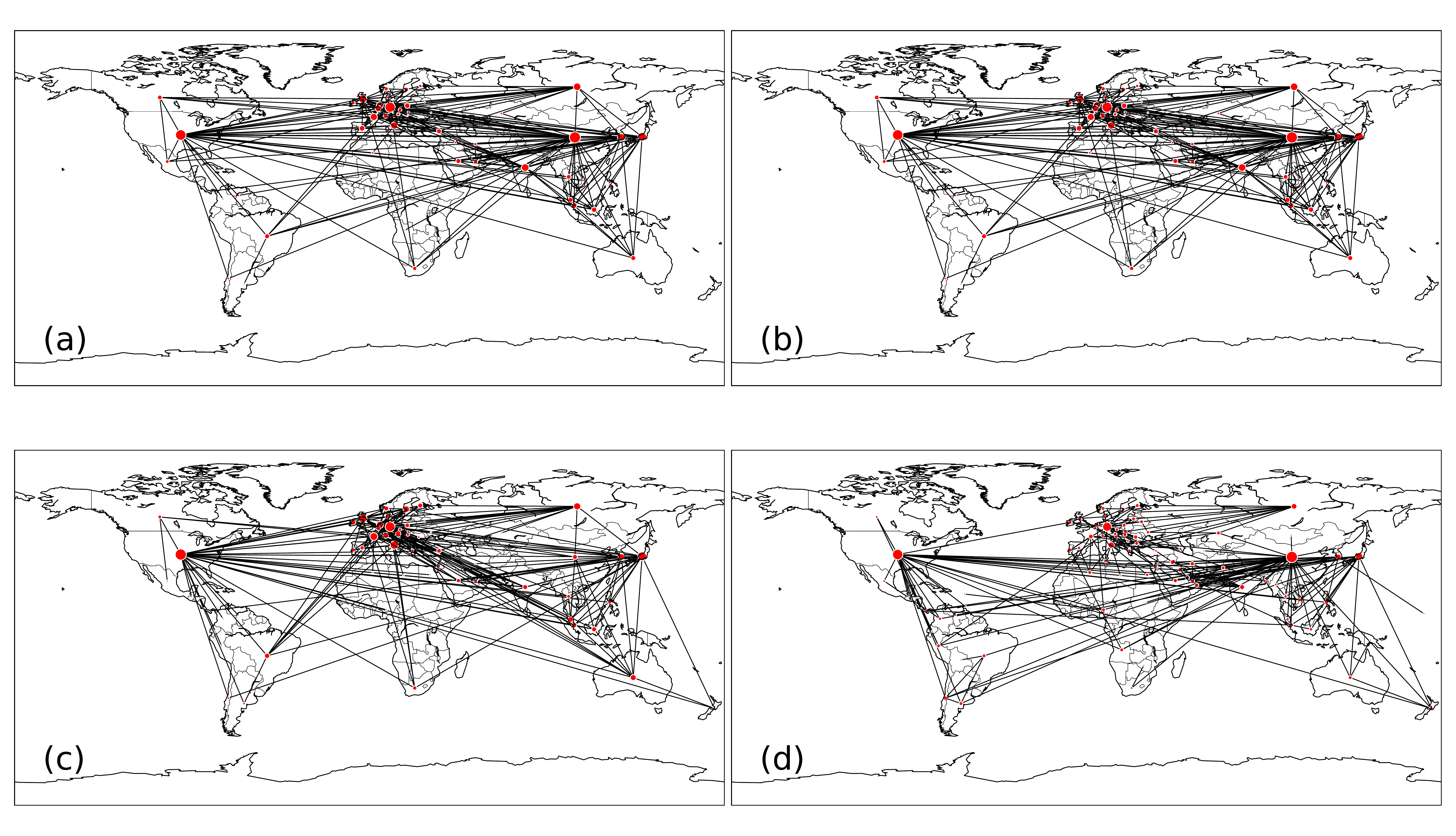}
\caption{Filtered backbones for the International Trade Networks network in year 2011. In each map, we display the top 200 heaviest connections for the original (a), disparity (b), \gloss (c) and ECM (d) networks. All the filtered backbones have been obtained using a $p$-value, $\widetilde{\gamma}$, equal to 0.05.}
\label{fig:trade-methods}
\end{figure*}

The case of the International Trade Network (in the year 2011) exhibits trends similar to US airports. In particular, as reported in Fig.~\ref{fig:trade-methods}, the disparity backbone and the original network (panels b and a) look very alike, having China (CHN) and USA playing the role of global juggernauts since they embody together the $32.5 \%$ of all connections. We also notice the role of global broker/middleman played by Europe as well as the presence of members of G8 as Russia (RUS) and Japan (JPN), together with some G20 members like India (IND), South Korea (KOR), Brazil (BRA), South Africa (ZAF), Australia (AUS) and Indonesia (IDN). 
However, the complete absence of connections either within or towards African countries (except for South Africa) looks quite unrealistic. 
The backbone obtained using GloSS, albeit resembling the original one, looks more like a star with Europe at its center, in line with its geographical, political and technological role. Unfortunately, in the map depicted by \gloss it is hard to discriminate any local relationship between neighbouring countries which surely exists due to their tight related historical development. 
As in the case of airports, the scenario depicted by ECM (panel d) is rather different from the previous two and is the least predictable from the original network. Some of the features captured by disparity and \gloss can still be found in the ECM backbone, like the prominent role of USA, China and Europe on the global checkerboard. Others are captured by ECM only, and can thus be considered the hallmarks of ECM itself. We observe the emergence of many more ``spheres of influence'' characterized by an interaction pattern which is stronger with neighbouring countries in close analogy with what observed for airports. For example, Russia loses its global role and becomes almost exclusively a partner of European countries. Brazil has more connections with other South American countries. African countries other than South Africa like Nigeria (NGA), Angola (AGO) and North African countries such as Morocco (MAR) and Egypt (EGY) appear. Australia becomes more pivotal in the South Pacific.
An unexpected trait highlighted by ECM is the brokering role between USA and China played by Middle Eastern countries like Saudi Arabia (SAU). Finally, Europe loses its role of global broker and becomes a more independent player.

\begin{figure*}[t]
\centering
\includegraphics[width=1.0\textwidth]{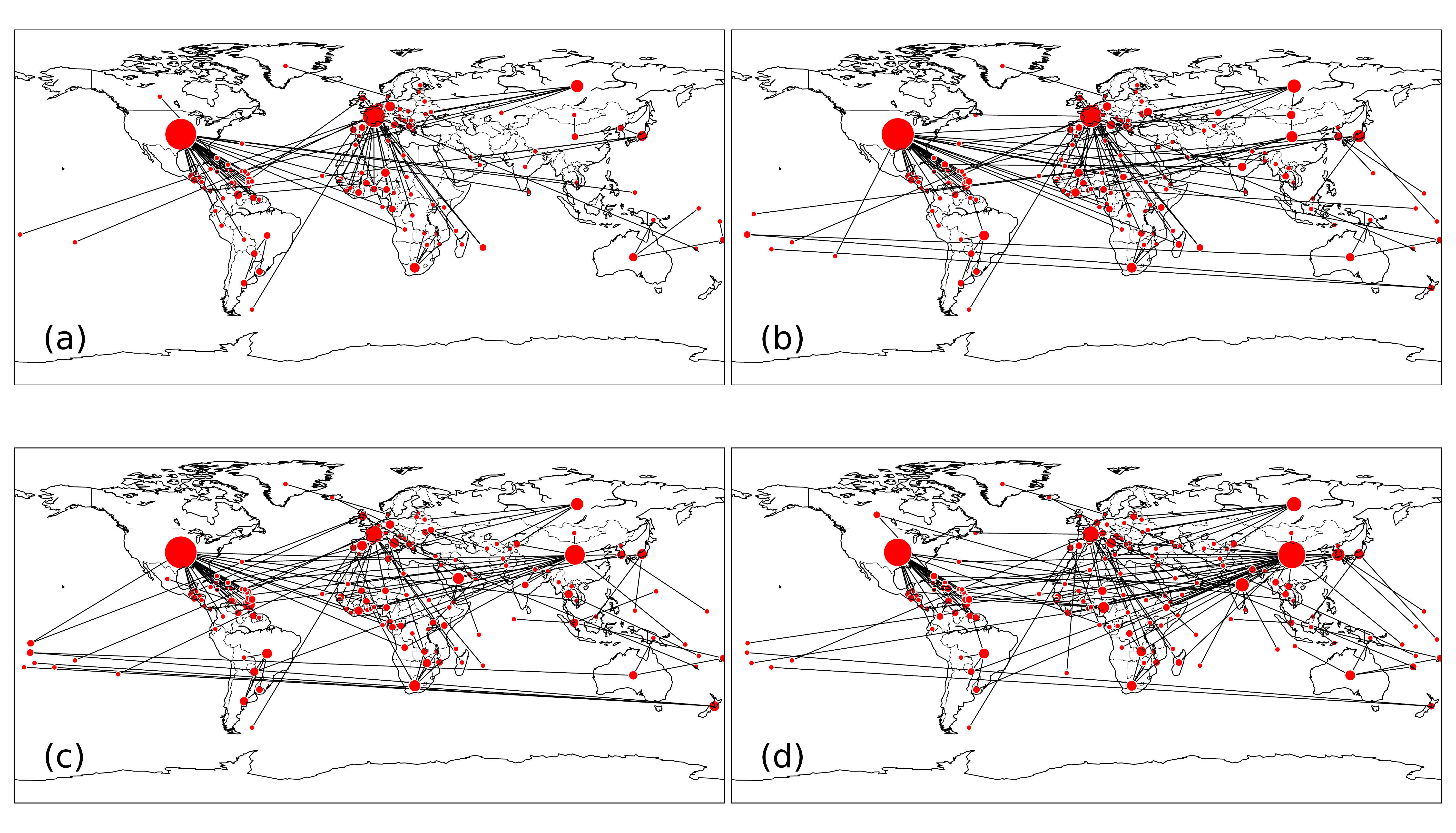}
\caption{Time evolution of the International Trade Network, filtered according to the local ECM filter. Top left: 1998; top right: 2002; bottom left: 2006; bottom right: 2011. Figures refer to a critical $p$-value of $ 10^{-6} $.} 
\label{fig:trade_evol_time_filtered}
\end{figure*}

Finally, we comment on the ability of ECM to identify relevant features \emph{per se}. As an illustrative example, we consider the time evolution of the International Trade Network in the period 1998--2011 displayed in Fig.~\ref{fig:trade_evol_time_filtered}. In 1998, we can distinguish basically six ``centers of influence''.
Two of them (namely USA and France (FRA)) act as global partners, while the other four, \ie Russia (RUS), South Africa (ZAF), Australia (AUS) and Japan (JPN), appear instead to play a more ``local'' role. As time passes, we notice the rise of some countries and the fall of others. For example, around year 2002 we notice the growth of China (CHN), South Korea (KOR) and India (IND). In 2006 France (FRA) has considerably lost its original influence while New Zealand (NZL) plays a prominent role among the Pacific islands compartment (though showing connections which are less relevant in terms of exchanged volumes); moreover, China still exhibits a startling development. Finally, in 2011 China and USA appear to be equally influential centers of trade. In particular, China gains several connections with the African countries, to the detriment of France and other European countries.

\section{Additional specifications of the method\label{sec:specifications}}
We now illustrate how our method can be extended to different ensembles of weighted networks. For the sake of brevity, we only provide the mathematical expressions and do not show explicit empirical analyses.

\subsection{Extension to directed networks}

Let us consider the set $\cal{W}$ of \emph{directed} weighted networks with $N$ nodes, each of which is described by a $N\times N$ weight matrix $\bm{W}$ that is not necessarily symmetric and has non-negative integer entries.
The constraints we impose are now the out-degree $k^\textrm{out}_i$, the in-degree $k^\textrm{in}_i$ (defined as the number of out-going and in-coming links of node $i$, respectively), the out-strength $s^\textrm{out}_i$ and the in-strength $s^\textrm{in}_i$ (defined as the total out-going  and in-coming weight of the links of node $i$, respectively).
If we denote empirical values by asterisks, enforcing these constraints on average means requiring
\begin{eqnarray}
&\langle \vec{k}^\textrm{ out}\rangle=\vec{k}^{\textrm{ out}\ast},\quad 
\langle \vec{k}^\textrm{ in}\rangle=\vec{k}^{\textrm{ in}\ast},&\\
&\langle \vec{s}^\textrm{ out}\rangle=\vec{s}^{\textrm{ out}\ast},\quad 
\langle \vec{s}^\textrm{ in}\rangle=\vec{s}^{\textrm{ in}\ast},&
\end{eqnarray}
which results in the Hamiltonian 
\begin{eqnarray}
&H&\left( \bm{W} \right|\vec{x}^\textrm{ out},\vec{x}^\textrm { in},\vec{y}^\textrm{ out},\vec{y}^\textrm{ in})=\\
&=& -\sum_{i=1}^N\sum_{j\ne i} \left[ \Theta(w_{ij})\ln(x^\textrm{out}_i x^\textrm{in}_j)+  w_{ij}\ln(y^\textrm{out}_iy^\textrm{in}_j) \right],\nonumber
\end{eqnarray}
where $x^\textrm{out}_i$, $x^\textrm{in}_i$, $y^\textrm{out}_i$, $y^\textrm{in}_i$ are Lagrange multipliers coupled to $k^\textrm{out}_i$, $k^\textrm{in}_i$, $s^\textrm{out}_i$, $s^\textrm{in}_i$ respectively.
A straightforward modification of the calculation we showed for the undirected case leads to the following results.
The graph probability that maximizes the Shannon-Gibbs entropy 
subject to the above constraints is
\begin{equation}
P(\bm{W}|\vec{x}^\textrm{ out},\vec{x}^\textrm { in},\vec{y}^\textrm{ out},\vec{y}^\textrm{ in}) 
=\prod_{i=1}^N \prod_{j\ne i} q_{ij}(w_{ij}),
\end{equation}
where 
\begin{equation}
q_{ij}(w)=\left\{\begin{array}{ll}1-p_{ij}&\textrm{ if}\quad w=0\\
p_{ij}\left(y^\textrm{out}_i y^\textrm{in}_j\right)^{w-1}(1-y^\textrm{out}_i y^\textrm{in}_j)&\textrm{ if}\quad w>0\end{array}\right.,\nonumber
\end{equation}
is the probability that the \emph{directed} link from node $i$ to node $j$ has weight $w$, and
\begin{equation}
p_{ij}\equiv1-q_{ij}(0)=\dfrac{x^\textrm{out}_i x^\textrm{in}_j y^\textrm{out}_i y^\textrm{in}_j}{1 - y^\textrm{out}_i y^\textrm{in}_j + x^\textrm{out}_i x^\textrm{in}_j y^\textrm{out}_i y^\textrm{in}_j}
\end{equation}
is the probability that a directed link from node $i$ to node $j$ exists, irrespective of its weight.

Given a real directed network $\bm{W}^\ast$, the values of the Lagrange multipliers are found by maximizing the log-likelihood
\begin{equation}
\ln P(\bm{W^{\ast}} | \vec{x}^\textrm{ out},\vec{x}^\textrm { in},\vec{y}^\textrm{ out},\vec{y}^\textrm{ in})
\end{equation}
or, equivalently, as the solution to the following $4N$ coupled equations:
\begin{eqnarray}
k^{\textrm{ out}\ast}_i&=&\sum_{j\ne i}\frac{x^\textrm{out}_i x^\textrm{in}_j y^\textrm{out}_i y^\textrm{in}_j}{1 - y^\textrm{out}_i y^\textrm{in}_j + x^\textrm{out}_i x^\textrm{in}_j y^\textrm{out}_i y^\textrm{in}_j}\nonumber\\
k^{\textrm{ in}\ast}_i&=&\sum_{j\ne i}\frac{x^\textrm{out}_j x^\textrm{in}_i y^\textrm{out}_j y^\textrm{in}_i}{1 - y^\textrm{out}_j y^\textrm{in}_i + x^\textrm{out}_j x^\textrm{in}_i y^\textrm{out}_j y^\textrm{in}_i}\nonumber\\
s_i^{\textrm{ out}\ast}&=&\sum_{j\ne i}\frac{x^\textrm{out}_i x^\textrm{in}_j y^\textrm{out}_i y^\textrm{in}_j}{(1 - y^\textrm{out}_i y^\textrm{in}_j + x^\textrm{out}_i x^\textrm{in}_j y^\textrm{out}_i y^\textrm{in}_j)(1-y^\textrm{out}_i y^\textrm{in}_j)}
\nonumber\\
s_i^{\textrm{ in}\ast}&=&\sum_{j\ne i}\frac{x^\textrm{out}_j x^\textrm{in}_i y^\textrm{out}_j y^\textrm{in}_i}{(1 - y^\textrm{out}_j y^\textrm{in}_i + x^\textrm{out}_j x^\textrm{in}_i y^\textrm{out}_j y^\textrm{in}_i)(1-y^\textrm{out}_j y^\textrm{in}_i)}\nonumber
\end{eqnarray}

Once the parameter values are found, the $p$-value for the weight $w_{ij}^*>0$ of the realized directed link from node $i$ to node $j$ reads
\begin{equation}
\gamma_{ij}^{\ast} \equiv \textrm{Prob}(w_{ij} \geq w_{ij}^{\ast})=
p^\ast_{ij}\left( y^{\textrm{out}\ast}_i y^{\textrm{in}\ast}_j \right)^{w_{ij}^{\ast}-1}.
\end{equation}
As before, the local filter proceeds by retaining only the links for which the $p$-value $\gamma_{ij}^{\ast}$ is smaller than a chosen critical value $\widetilde{\gamma}$.
The global filter would employ a similar criterion based on the probability mass function, rather than on the cumulative probability function, but this is expected to lead to poorer results, as already discussed for the undirected case.

\subsection{Extension to bipartite networks}
We then assume that $\bm{W}^\ast$ is a \emph{bipartite}, undirected, weighted network, and $\cal{W}$ the corresponding ensemble.
Each network in the ensemble has two layers, one with $N_1$ nodes and one with $N_2$ nodes. 
Links are only allowed across layers, not within them.
For each node $i$, one can still define the degree $k_i$ and strength $s_i$ as for an ordinary (\ie unipartite) undirected graph.
The main difference with respect to the unipartite case is the fact that all graphs $\bm{W}$ that do not have a bipartite structure are excluded from $\cal{W}$ and from the calculations.

There is, however, a trick that allows us to map (exactly) the ensemble of bipartite undirected graphs to the ensemble of unipartite directed graphs considered above.
The trick consists in assigning an arbitrary but common direction (say, from layer 1 to layer 2) to all the links in the original bipartite network $\bm{W}^*$. Then, the resulting directed network can be treated as a unipartite one with $N=N_1+N_2$ nodes and the procedure described above for directed networks can be applied. 
At the end, the direction of the links that are retained by the filter is simply discarded, and one correctly obtains the irreducible backbone of the original bipartite undirected network.

The above mapping between a bipartite undirected graph and a unipartite directed graph (and the corresponding null models) is exact because, after assigning a direction to the links, all nodes in (say) layer 1 have $k^{in}_i=0$ and $k^{out}_i>0$, while all nodes in (say) layer 2 have $k^{out}_i=0$ and $k^{in}_i>0$ (if we had $k^{in}_i=0$ and $k^{out}_i=0$, node $i$ would be disconneted from all other nodes, and we would have discarded it). Similar conditions hold for the in- and out-strenghts.
If we now apply the null model for weighted directed unipartite networks described in the previous subsection, the zero in- and out-degrees (and strengths) will be kept to zero, as there is no other way to enforce that their expected value is zero.
This ensures that the nodes in each layer do not receive connections from other nodes in the same layer, so that the bipartite structure is preserved in the null model as desired.

\section{Conclusions\label{sec:conclusions}}

The ever-increasing availability of `big data' has spurred the use of networks as a powerful way to capture the relevant features of complex systems. However, when the flood of information becomes overwhelming, the advantages of a network representation tend to fade away and the possibility to discriminate the essential structure of the system (\ie its backbone) deteriorates considerably. 
To preserve sparsity and non-redundancy of networks, several filtering techniques have been developed so far. 

At the same time, recent improvements in network reconstruction techniques have emphasized that many structural features of real-world networks can be reliably estimated from the knowledge of the local node-specific topological properties, namely the degrees and strengths of nodes. 
This means that the truly dyadic relationships, \ie those that are irreducible to node properties, may be hidden amidst a majority of redundant ones.
In particular, recent results have shown that the ``first-order approximation'' for many networks with heterogeneous nodes is the ECM, while any other feature not directly encapsulated in the size of nodes (like higher-than-expected preference for specific connections, dependence on geographic or other distances, presence of communities and motifs, etc.) is expected to be immediately visible at the next order. 

Based on the above considerations, we have introduced a method that filters out the first-order local effects embodied in the strength and degree sequence, thus highlighting the truly dyadic (and higher-order) patterns relating nodes to each other. 
We found that, while before applying the filter many networks display similar properties (precisely because their first-order structure is well approximated by the ECM), after applying the filter they show significant differences, presumably because higher-order features arise as network-specific effects.

Importantly, since nodes and degrees are the maximal set of local node-specific properties that can be defined in any weighted network, our approach is guaranteed to identify the connections that are by construction impossible to infer on the basis of node-specific properties alone and that cannot be recovered by any network reconstruction method based only on local node properties.

The comparison of the performances of the ECM, disparity and \gloss methods shows that the ECM filter outperforms its competitors. We have also examined the structural differences between the backbones retrieved from the global and local implementations of our filter and the role of the order of filtering and aggregating in multiplex networks. 
We have applied the ECM filter to the analysis of several empirical datasets and illustrated how successfully it extracts relevant hidden features like the hub-and-spoke structure of the US airport network and the evolution in time of the most relevant ``spheres of influence'' across world trade. 

Finally, we have shown that the ECM filter can be applied to different kinds of weighted networks (e.g. undirected, directed, bipartite) and therefore constitutes a valuable tool for the analysis of any networked system where the excess of information hinders the identification of the essential backbone of interactions.

\section*{Acknowledgments}
We acknowledge support from the Dutch Econophysics Foundation (Stichting Econophysics, Leiden, the Netherlands) and the Swiss National Science Foundation under Grant No. CRSII2\_147609.

%
%

\pagebreak
\widetext
\begin{center}
\textbf{\large Supplementary Materials for the manuscript entitled: \\``Irreducible network backbones: unbiased graph filtering via maximum entropy''}
\end{center}
\setcounter{equation}{0}
\setcounter{figure}{0}
\setcounter{table}{0}
\setcounter{page}{1}
\setcounter{section}{0}
\makeatletter
\renewcommand{\thetable}{TS\arabic{table}}
\renewcommand{\thefigure}{S\arabic{figure}}
\renewcommand{\thesection}{S\Roman{section}} 
\renewcommand{\thesubsection}{\thesection.\roman{subsection}}

\renewcommand{\arraystretch}{1.15}

\section{Datasets}

The present section contains the results that have not been displayed in the main text grouped by datasets.

\subsection{International Time-varying Trade Network}

\begin{figure*}[hb!]
\centering
\includegraphics[width=1.0\columnwidth]{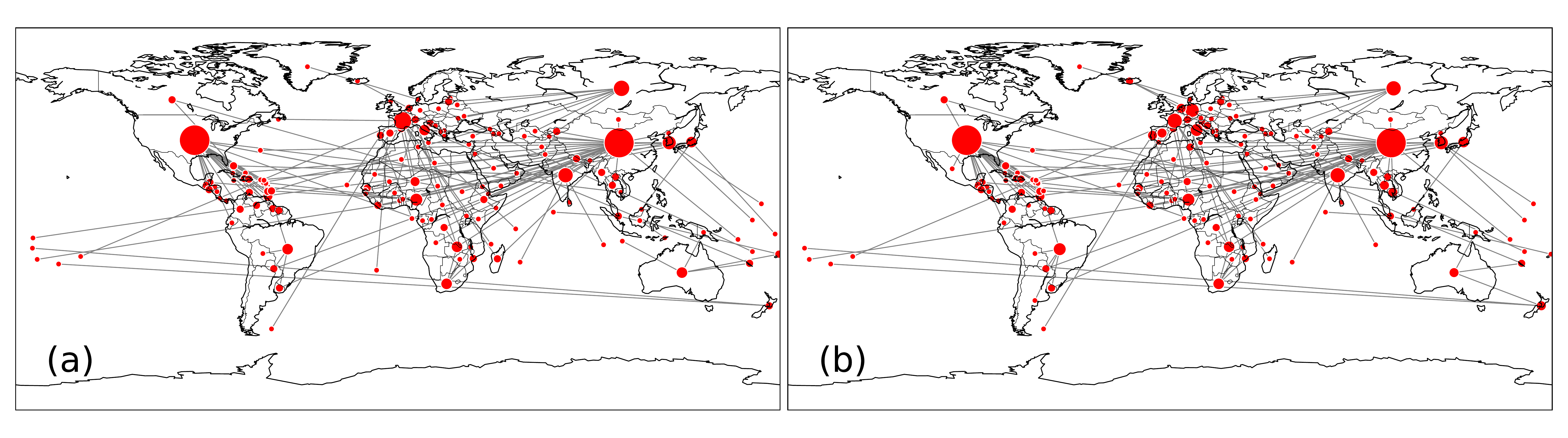}
\caption{2011 World Trade Networks filtered using local (a) and global (b) ECM filters. The local network is obtained using $\widetilde{\gamma} = 10^{-6}$. The number of links, $L$, in both graphs is 149. The size of the nodes is proportional to their degree.}
\label{fig:maps_local_global}
\end{figure*}

Considering the trading volumes between countries in the period between the years 1998 and 2011 we can build a time-varying network where each time snapshot corresponds to a given year. A node indicates a country and the weight of a link denotes the gross trade volume between two countries. In Fig.~\ref{fig:maps_local_global} we show the filtered graphs for the year 2011. Panel (a) is the local filter case obtained considering $p$-value, $\widetilde{\gamma}$, equal to $10^{-6}$. Panel (b) is the global case obtained choosing the first $ L^{\prime} $ least likely links such that the number of edges in the two networks is the same. At a glance, we see that most of the significant connections are shared by both graphs.

%
%
%

In Tab.~\ref{tab:tvg_trade_1998_top_20_signific} we report the list of the twenty most significant links according to local ECM, global ECM, \gloss and Disparity filters for year 1998 \cite{radicchi-pre-2011}. The names of the countries are represented using the \textbf{ISO 3166-1 alpha-3} standard encoding \cite{iso3-nations-wikipedia}. Since we are looking at the first twenty most significant links out of about 200, it is reasonable that the four lists are pretty much similar. Interestingly, the most significant connection in \gloss and Disparity (\ie that between Canada (CAN) and USA (USA)) is not even present among the most significative of ECM filters. Disparity, instead, fails to identify the links between France (FRA) and French Polynesia (PYF) and also between Belgium-Luxemburg (BLX) and Central African Republic (CAF) to give an example. More in general, DF tends to assign an eccessive relevance to USA placing fifteen out of twenty connections with USA in the list while for the other methods such number drops to just half of the connections. Finally, it is worth mentioning that with the sole exception of DF, we are not able to find trace of China in 1998 table in accordance with the not predominant role played by such country at that time.
%
%
%
%
\begin{table}[ht!]
\centering
 \begin{small}
 \begin{tabular}{||c|c|c|c|c||} 
 \hline
 Rank & Local filter & Global filter & GloSS & Disparity\\ [0.5ex] 
 \hline\hline
 1 & DNK GRL & DNK GRL & CAN USA & CAN USA \\ \hline
 2 & JAM USA & JAM USA & DNK GRL & MEX USA \\ \hline
 3 & HND USA & HND USA & TCA USA & BLR RUS \\ \hline
 4 & HTI USA & BLR RUS & COM FRA & DOM USA \\ \hline
 5 & VGB RUS & DOM USA & HTI USA & HND USA \\ \hline
 6 & BLR RUS & HTI USA & VGB RUS & AUT DEU \\ \hline
 7 & FRA PYF & VGB RUS & BLX CAF & CRI USA \\ \hline
 8 & DOM USA & FRA PYF & JAM USA & CZE DEU \\ \hline
 9 & BLX CAF & CRI USA & FRA PYF & JAM USA \\ \hline
 10 & FRA MDG & GAB USA & CPV PRT & VGB RUS \\ \hline
 11 & GAB USA & FRA MDG & AND ESP & JPN USA \\ \hline
 12 & AND ESP & AND ESP & FRA MDG & COL USA \\ \hline
 13 & CRI USA & BLX CAF & HND USA & HTI USA \\ \hline
 14 & ALB ITA & ALB ITA & GRD USA & GTM USA \\ \hline
 15 & TCA USA & GTM USA & ALB ITA & GAB USA \\ \hline
 16 & CPV PRT & NIC USA & BLR RUS & IRL GBR \\ \hline
 17 & COM FRA & BHS USA & GAB USA & CHN USA \\ \hline
 18 & NIC USA & ANT VEN & FRA WLF & ECU USA \\ \hline
 19 & BHS USA & CPV PRT & CRI USA & ISR USA \\ \hline
 20 & GTM USA & TTO USA & DOM USA & TCA USA \\ \hline
\end{tabular}
\end{small}
\caption{List of the twenty most significant links in the International Trade Network for year 1998 according to the local and global ECM filters, GloSS and Disparity Filter.}
\label{tab:tvg_trade_1998_top_20_signific}
\end{table}
In Tab.~\ref{tab:tvg_trade_2011_top_20_signific_all} we list the same kind of information displayed in Tab.~\ref{tab:tvg_trade_1998_top_20_signific} but for year 2011. At first glance, something catches our attention: the presence of the following ``\emph{bizarre}'' connections: Antigua and Barbuda (ATG) and Nigeria (NGA), Algeria (DZA) and Saint Kitts and Nevis (KNA), Barbados (BRB) and Nigeria, Turks and Caicos Islands (TCA) and USA. All methods indicate those connections as relevant. However, a deeper analysis of the data revealed the presence of errors in the records of trade volumes among such countries. Some of the endpoints of these anomalous connections are very small Caribbean or Pacific countries, showing sheer trade volumes (for certain commodities) higher than those between China and USA for example. In the light of these findings, filtering can be thought of not exclusively as a way to recognize relevant connections, but also as a method to validate them. Finally, contrary to the Disparity Filter, ECM and \gloss identify additional wrong entries in the dataset such as: Bermuda (BMU) and South Korea, South Korea and Liberia (LBR), Nigeria and Niue (NIU) and Cocos (Keeling) Islands (CCK) and India (IND). We have checked for the presence of such anomalous connections across all our time-varying data, and we have found that such mistakes are present only in the years 2009, 2010 and 2011. Among other noticeable connections spotted by ECM filter, instead, we find: South Africa (ZAF) and Zimbawe (ZWE), Denmark (DNK) and Greenland (GRL), China (CHN) and Mongolia (MNG), Albania (ALB) and Italy (ITA) just to cite a few.

%
%
%
\begin{table}[ht!]
\centering
 \begin{small}
 \begin{tabular}{||c|c|c|c|c||} 
 \hline
 Rank & Local filter & Global filter & GloSS & Disparity\\ [0.5ex] 
 \hline\hline
 1 & {\cellcolor{green!25}} ATG NGA & {\cellcolor{green!25}} ATG NGA & {\cellcolor{green!25}} ATG NGA & {\cellcolor{green!25}} ATG NGA \\ \hline
 2 & CHN PRK & CHN PRK & CHN PRK & MEX USA \\ \hline
 3 & {\cellcolor{green!25}} DZA KNA & {\cellcolor{green!25}} DZA KNA & {\cellcolor{orange!25}} NGA NIU & CAN USA \\ \hline
 4 & NPL IND & NPL IND & BRA LCA & {\cellcolor{green!25}} TCA USA \\ \hline
 5 & {\cellcolor{green!25}} BRB NGA & {\cellcolor{green!25}} BRB NGA & {\cellcolor{green!25}} DZA KNA & NPL IND \\ \hline
 6 & {\cellcolor{green!25}} TCA USA & {\cellcolor{green!25}} TCA USA & NPL IND & {\cellcolor{green!25}} BRB NGA \\ \hline
 7 & AND ESP & AND ESP & AND ESP & BLR RUS \\ \hline
 8 & ZAF ZWE & ZAF ZWE & {\cellcolor{green!25}} BRB NGA & DOM USA \\ \hline
 9 & BRA LCA & BRA LCA & {\cellcolor{orange!25}} CCK IND & TCD USA \\ \hline
 10 & DOM USA & DOM USA & DNK GRL & AND ESP \\ \hline
 11 & DNK GRL & ABW USA & TCD USA & AUT DEU \\ \hline
 12 & TCD USA & TCD USA & ZAF ZWE & HND USA \\ \hline
 13 & ABW USA & DNK GRL & BTN IND & ABW USA \\ \hline
 14 & ALB ITA & HND USA & DOM USA & GTM USA \\ \hline
 15 & HND USA & ALB ITA & COM FRA & CRI USA \\ \hline
 16 & JAM USA & CHN SDN & PRT STP & ALB ITA \\ \hline
 17 & CHN SDN & JAM USA & ALB ITA & CZE DEU \\ \hline
 18 & CHN MNG & CHN MNG & ABW USA & BTN IND \\ \hline
 19 & BTN IND & {\cellcolor{cyan!25}} KOR LBR & COK NZL & COL USA \\ \hline
 20 & {\cellcolor{cyan!25}} BMU KOR & BTN IND & JAM USA & SLV USA \\ \hline
\end{tabular}
\end{small}
\caption{List of the twenty most significant links in the International Trade Network for year 2011 according to the local and global ECM filters, GloSS and Disparity filter. Green cells correspond to connections displaying false volumes. Blue (orange) cells correspond to erroneous connections identified only by ECM (\gloss) filter.}
\label{tab:tvg_trade_2011_top_20_signific_all}
\end{table}

The scenario becomes more interesting by looking at Tab.~\ref{tab:tvg_trade_2011_top_20_heaviest}, i.e. the list of the twenty heaviest links according to the global and local methods. Here, in fact, we can see how the link between China (CHN) and USA (USA), which is the heaviest in the original network, has disappeared from both filtered networks. Conversely, the relation between Russian Federation (RUS) and Ukraine (UKR) clearly emerges in the filtered networks as one of the most important ones. Another curious feature is the vanishing of Germany (DEU) from the column of local filter albeit it appears in ten out of twenty positions available in the original networks. Finally, we observe the presence of a link between Italy (ITA) and Libya (LBY) which have a strong historic and economic relation due to the past role of Libya as one of the colonies of Italy during the beginning of the 20$^{\text{th}}$ century.

%
%
%
\begin{table}[ht!]
\centering
 \begin{small}
 \begin{tabular}{||c|c|c|c||} 
 \hline
 Rank & Original & Local & Global \\ [0.5ex] 
 \hline\hline
 1 & CHN USA & RUS UKR & CAN USA  \\ \hline
 2 & CHN JPN & USA VEN & MEX USA  \\ \hline
 3 & CAN USA & BLR RUS & AUT DEU  \\ \hline
 4 & MEX USA & COL USA & RUS UKR  \\ \hline
 5 & CHN KOR & AGO CHN & USA VEN  \\ \hline
 6 & FRA DEU & JPN PAN & DEU HUN  \\ \hline
 7 & CHN DEU & CHN OMN & BLR RUS  \\ \hline
 8 & DEU NLD & ECU USA & COL USA  \\ \hline
 9 & JPN USA & LTU RUS & ARG BRA  \\ \hline
 10 & DEU ITA & CRI USA & JPN QAT  \\ \hline
 11 & BLX NLD & AZE ITA & AGO CHN  \\ \hline
 12 & DEU GBR & GTM USA & PRT ESP  \\ \hline
 13 & BLX DEU & CHN SDN & JPN PAN  \\ \hline
 14 & DEU USA & FRA TUN & CHN AMN  \\ \hline
 15 & DEU CHE & HND USA & ECU USA  \\ \hline
 16 & AUS CHN & KOR LBR & LTU RUS  \\ \hline
 17 & JPN KOR & TTO USA & AUS NZL  \\ \hline
 18 & AUT DEU & KOR MHL & CRI USA  \\ \hline
 19 & BLX FRA & ITA LBY & DEU SVN  \\ \hline
 20 & DEU POL & CHN MNG & AZE ITA  \\ \hline
\end{tabular}
\end{small}
\caption{List of the twenty heaviest links in the International Trade Network (2011) in the original network, and according to the local and global ECM filters.}
\label{tab:tvg_trade_2011_top_20_heaviest}
\end{table}

\newpage

\subsection{2011 International Multiplex Trade Network}

\begin{figure*}[h!]
\centering
\includegraphics[width=1.0\columnwidth]{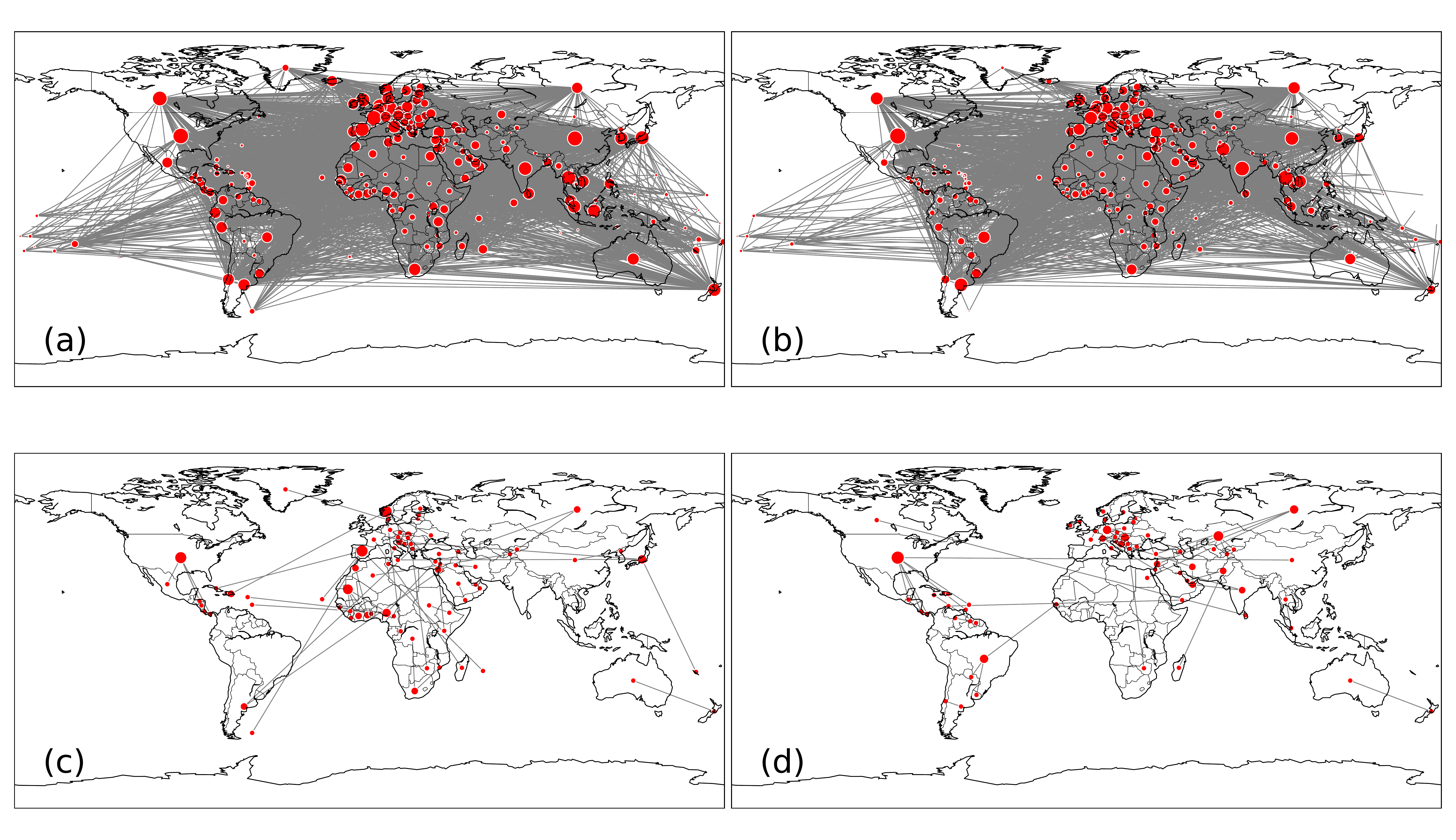}
\caption{Original and ECM filtered trade multiplex of fish and cereals. Panels a-c refer to the original (a) and local ECM filter (c) fish and crustaceans commodity. Panels b-d account for cereals, instead. The original networks are built using the 2011 data, and have been filtered considering $\widetilde{\gamma} = 10^{-5}$.}
\label{fig:maps_local_BACI2011_comm03_comm10}
\end{figure*}

\begin{figure*}[h!]
\centering
\includegraphics[width=1.0\columnwidth]{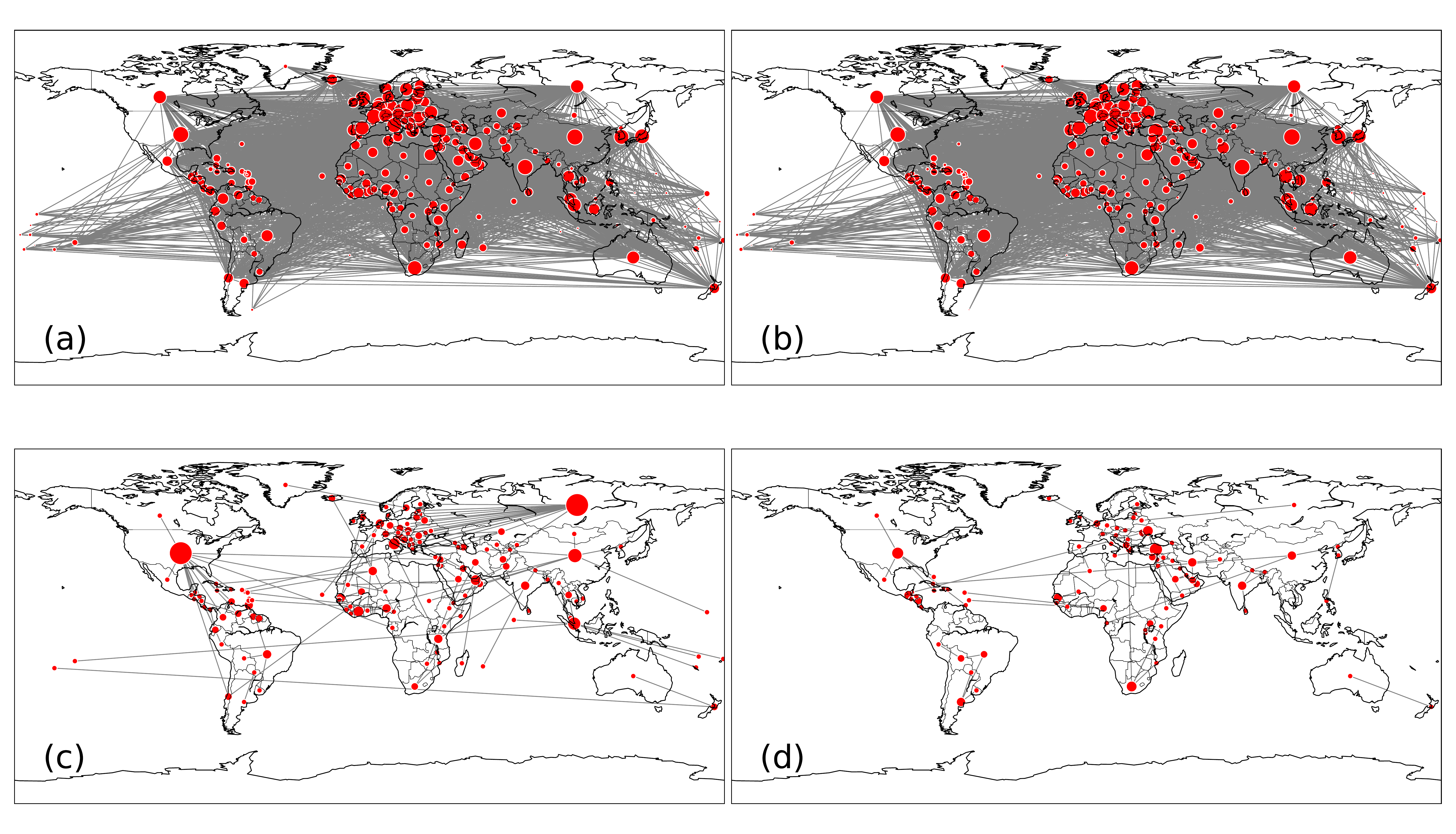}
\caption{Original (panels a-b) and ECM filtered (panels c-d) trade multiplex of fuels and oils (panels a-c) and iron and steel (panels b-d). The original networks are built using the 2011 data, and have been filtered considering $\widetilde{\gamma} = 10^{-5}$.}
\label{fig:maps_local_BACI2011_comm27_comm72}
\end{figure*}

\begin{figure*}[h!]
\centering
\includegraphics[width=1.0\columnwidth]{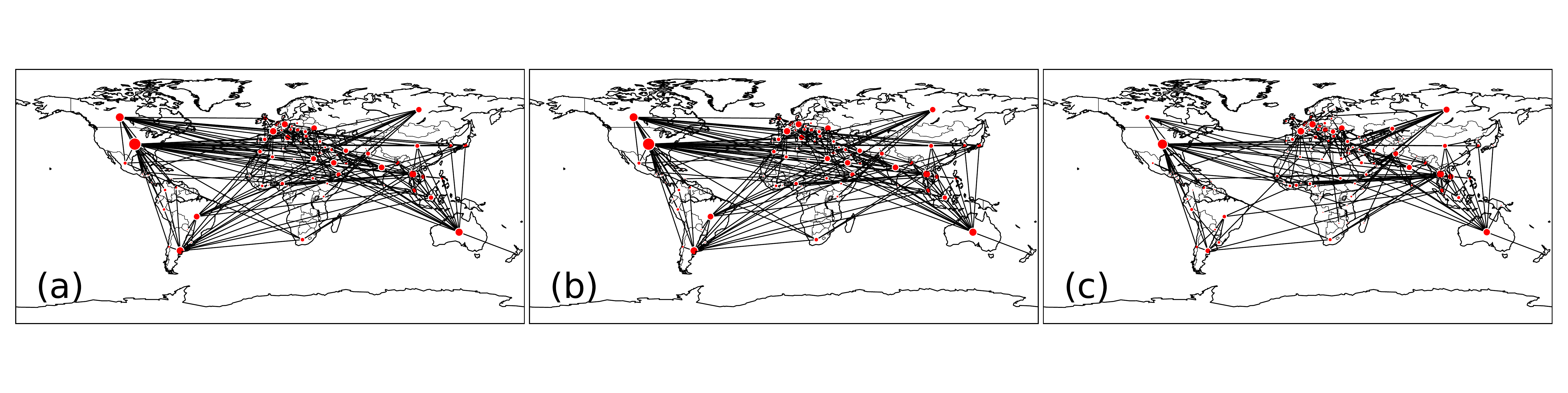}
\caption{Visual representation of the 200 heaviest links in the original network (a), ECM global (b) and local (c) filters for the 2011 Trade in cereals. The local filtering is obtained using $\widetilde{\gamma} = 0.05$, the global one is constructed such that $L_{GLOB} = L_{LOC}$.}
\label{fig:maps_comm10_heaviest}
\end{figure*}

The results displayed in Figures \ref{fig:maps_local_BACI2011_comm03_comm10} - \ref{fig:maps_comm10_heaviest} show that the ECM filter can be useful to detect patterns in the International Trade Multiplex, namely the multi-layer network where each node denotes a country and each layer represents the trade in a given commodity. Indeed, the original layers do not exhibit any evident difference among each other, due to the large density of this disaggregated representation; the filtered ones provide, instead, some relevant information, such as the appearance of Norway and Russia as hubs, respectively in the trade in fish and fuels/oil. 

\newpage

\newpage

\subsection{Airports}

\begin{figure}[h!]
\centering
\includegraphics[width=0.85\columnwidth]{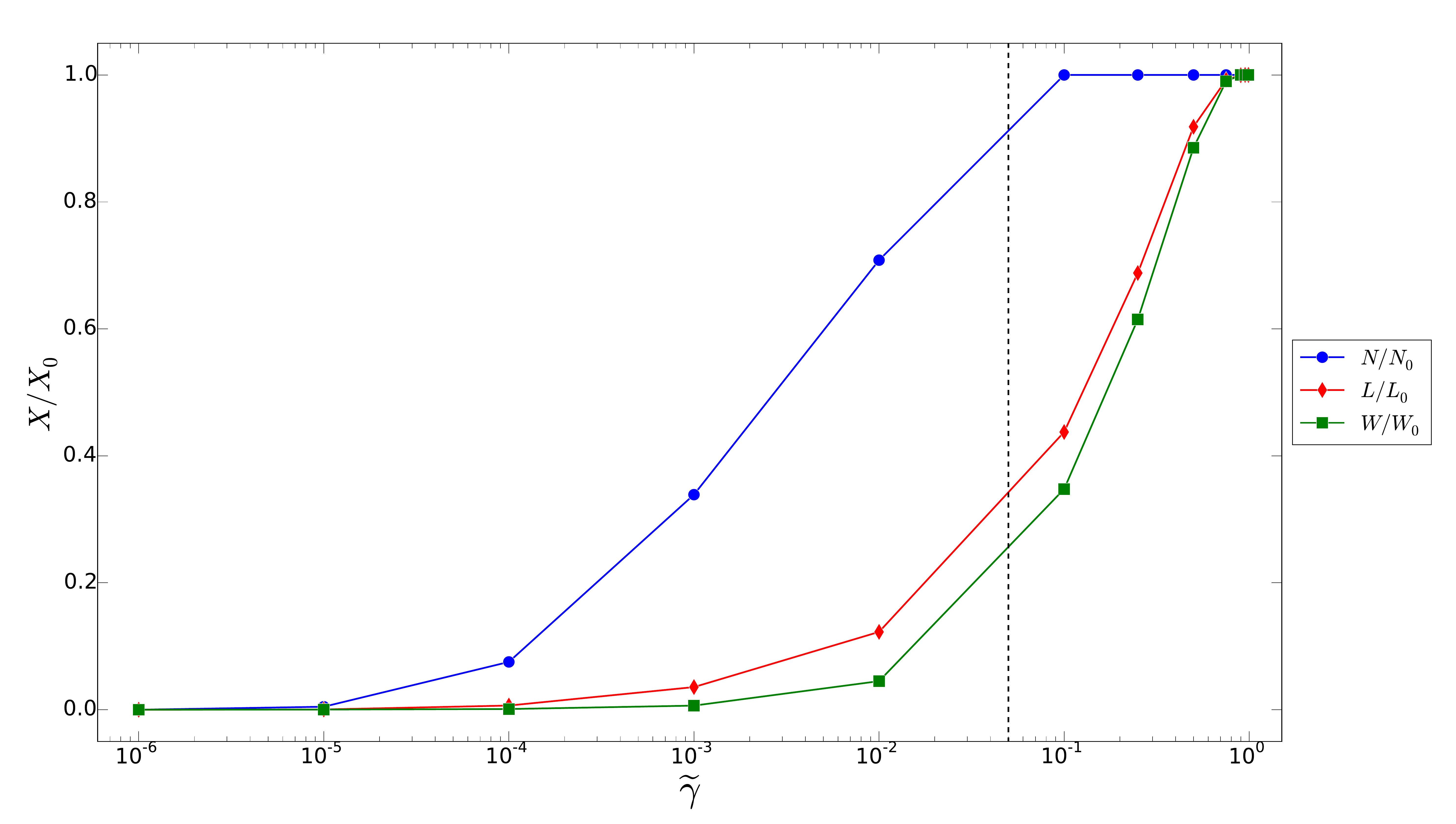}
\caption{Effect of filtering on the US-airport dataset. We report the fraction of the number of nodes $N/N_0$ (blue dots), edges $L/L_0$ (red diamonds) and total weight $W/W_0$ (green squares) as a function of the $p$-value $\widetilde{\gamma}$. The vertical line corresponds to $\widetilde{\gamma} = 0.05$.}
\label{fig:airport_filter_pval}
\end{figure}

\begin{figure*}[h!]
\centering
\includegraphics[width=1.0\columnwidth]{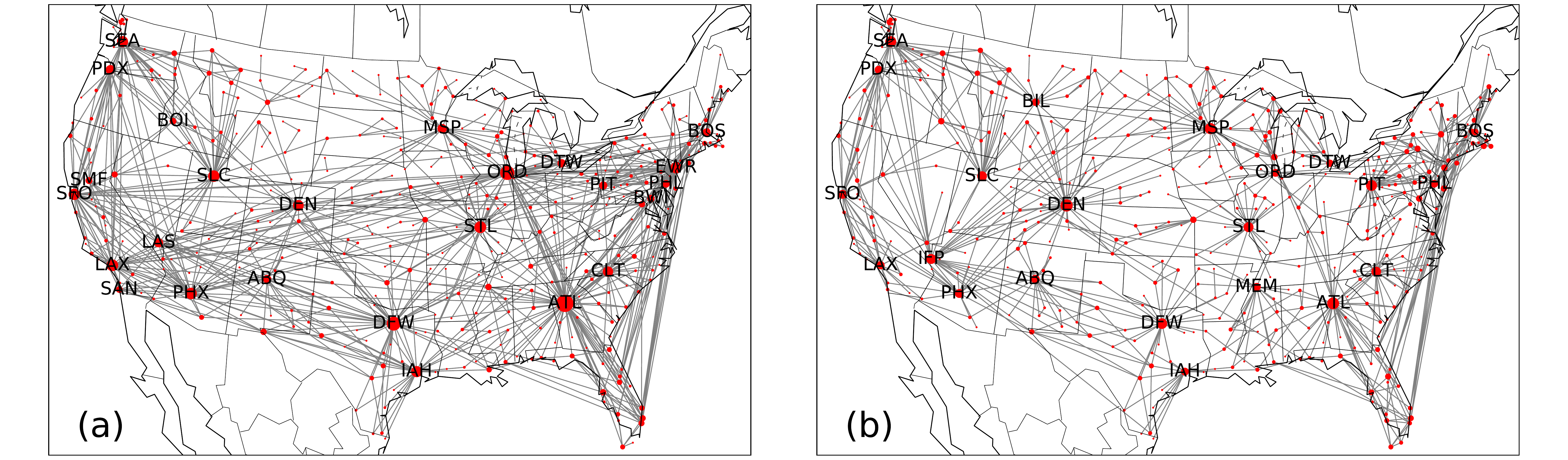}
\caption{Graphs obtained after filtering according to the global (left) and local (right) ECM method, for the US Airport Network. In the local filter, results refer to $\widetilde{\gamma} = 0.05$; in the global one, we choose the most unlikely links such that the number of edges in the two panels is the same (764).}
\label{fig:airport_local_global_alllinks}
\end{figure*}

In the US airport network, the first result reported in Fig.~\ref{fig:airport_filter_pval} is the behaviour of the fractions of nodes, edges and total weight as a function of the $p$-value. As we can see, for $\widetilde{\gamma} = 0.05$ the filter is able to remove around 70\% of the connections while retaining about 20\% of the total information (i.e. total weight). Analogously to Fig.~3 of main text, in Fig.~\ref{fig:airport_local_global_alllinks} we display the filtered networks with all their links. Despite being more ``noisy'', the difference between the structures of local and global networks remains clearly distinguishable. In particular, we observe the persistence of many links among principal aiports due to their heavy weights. The absence of such links in the local network enables the emergence of the hub-and-spoke structure mentioned in the main text.

%
%
%
%

%
%
%

In addition to the visual comparison between local and global filters, we propose the comparison between the twenty most significant edges (Tab.~\ref{tab:airports_global_local_20_significative}). The most striking fact about the names listed in this table is that they correspond mainly to very small towns. The only exception is link number 17 in the global network corresponding to the connection between Miami and Key West which is very important, among many factors, for tourism. This is completely different from what can be seen in Tab.~II of the main text, where all the connections listed are among main airports with New York and Los Angeles playing prominent roles in both the global and the local network.

%
%
%
\begin{table}[h!]
\centering
 \begin{tabular}{||c | c | c||} 
 \hline
 Rank & Local filter & Global filter \\ [0.5ex] 
 \hline\hline
 1 & Nantucket - Hyannis & Nantucket - Hyannis \\ 
 \hline
 2 & Bedford - Trenton & Bedford - Trenton \\
 \hline
 3 & Fort Dodge - Mason City & Fort Dodge - Mason City \\
 \hline
 4 & Alpena - Sault Ste Marie & Alpena - Sault Ste Marie \\
 \hline
 5 & Devils Lake - Jamestown & Lewiston - Pullman \\  
 \hline
 6 & Hot Springs - Harrison & Spokane - Seattle \\ 
 \hline
 7 & Denver - Bullhead City & Eau Claire - Rhinelander \\
 \hline
 8 & Kingman - Prescott & Devils Lake - Jamestown \\
 \hline
 9 & Brookings - Huron & Hancock - Marquette \\
 \hline
 10 & Melbourne - Oshkosh & Hot Springs - Harrison \\ 
 \hline
 11 & Hancock - Marquette & Columbus Starkville WestPt - Tupelo \\
 \hline
 12 & El Dorado - Jonesboro & Springfield - Quincy \\
 \hline
 13 & Eau Claire - Rhinelander & Grand Rapids - Saint Cloud \\
 \hline
 14 & Havre - Lewistown & Kingman - Prescott \\ 
 \hline
 15 & Grand Rapids - Saint Cloud & Friday Harbor - Lopez Island \\
 \hline
 16 & Clovis - Hobbs & Idaho Falls - Pocatello \\
 \hline
 17 & Riverton - Worland & Key West - Miami \\ 
 \hline
 18 & Lewiston - Pullman & Brookings - Huron \\
 \hline
 19 & North Platte - Norfolk & El Dorado - Jonesboro \\
 \hline
 20 & Manhattan - Salina & Melbourne - Oshkosh \\
 \hline
\end{tabular}
\caption{List of the 20 most significant links in the US Airport Network according to the local and global ECM filters. The local network is filtered considering $\widetilde{\gamma} = 0.05$.}
\label{tab:airports_global_local_20_significative}
\end{table}

\subsection{Florida Bay Food Web}

The analysis of the filtering power of local ECM on the Florida Bay dry dataset on the usual three topological indicators displays a behaviour not very different from the other cases. However, for $\widetilde{\gamma} = 0.05$ we observe a slightly higher value of $\tfrac{W}{W_0}$ than for airports accompanied by a steeper decrease of if for lower values of $\widetilde{\gamma}$.

%
%
%
\begin{figure}[h!]
\centering
\includegraphics[width=0.85\columnwidth]{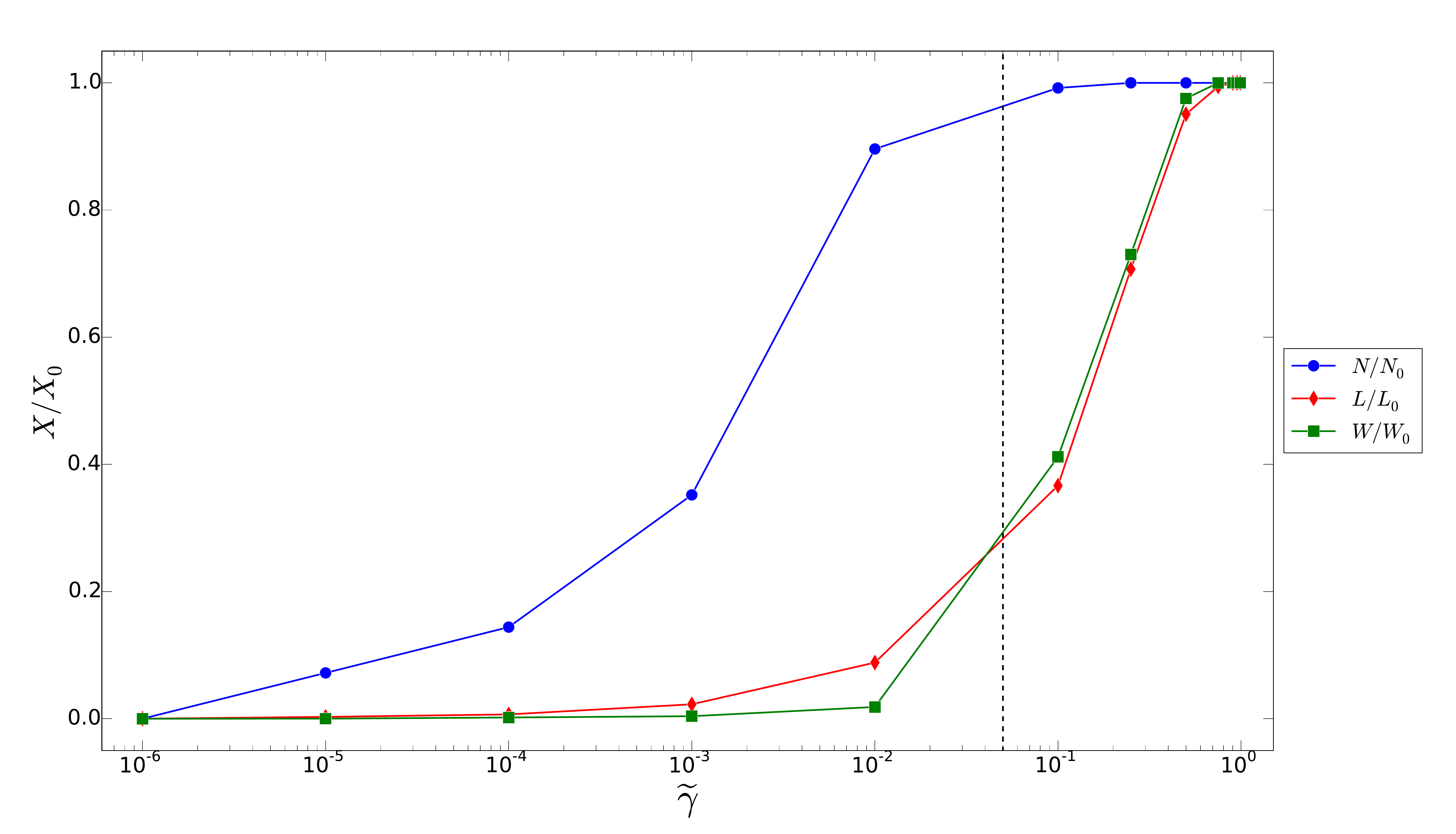}
\caption{Effect of filtering on the Florida Bay food web dataset. We report the fraction of the number of nodes $N/N_0$ (blue dots), edges $L/L_0$ (red diamonds) and total weight $W/W_0$ (green squares) as a function of the $p$-value $\widetilde{\gamma}$. The vertical line corresponds to $\widetilde{\gamma} = 0.05$.}
\label{fig:foodweb_filter_pval}
\end{figure}

\subsection{Star Wars}

\begin{figure}[h!]
\centering
\includegraphics[width=0.85\columnwidth]{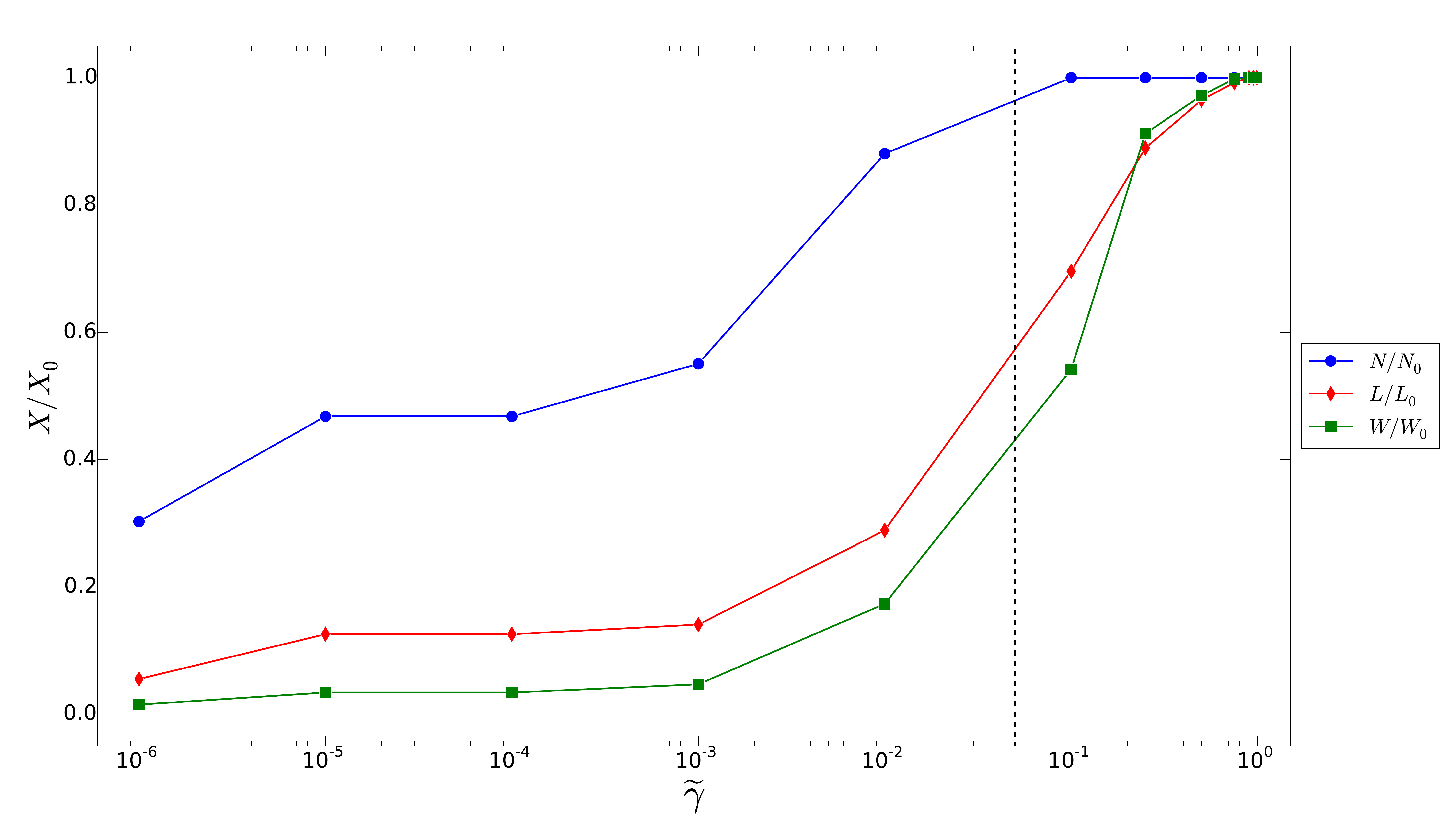}
\caption{Effect of filtering on the Star Wars interactions network. We report the fraction of the number of nodes $N/N_0$ (blue dots), edges $L/L_0$ (red diamonds) and total weight $W/W_0$ (green squares) as a function of the $p$-value $\widetilde{\gamma}$. The vertical line corresponds to $\widetilde{\gamma} = 0.05$.}
\label{fig:starwars_filter_pval}
\end{figure}

\begin{figure*}[h!]
\centering
\includegraphics[width=0.48\columnwidth]{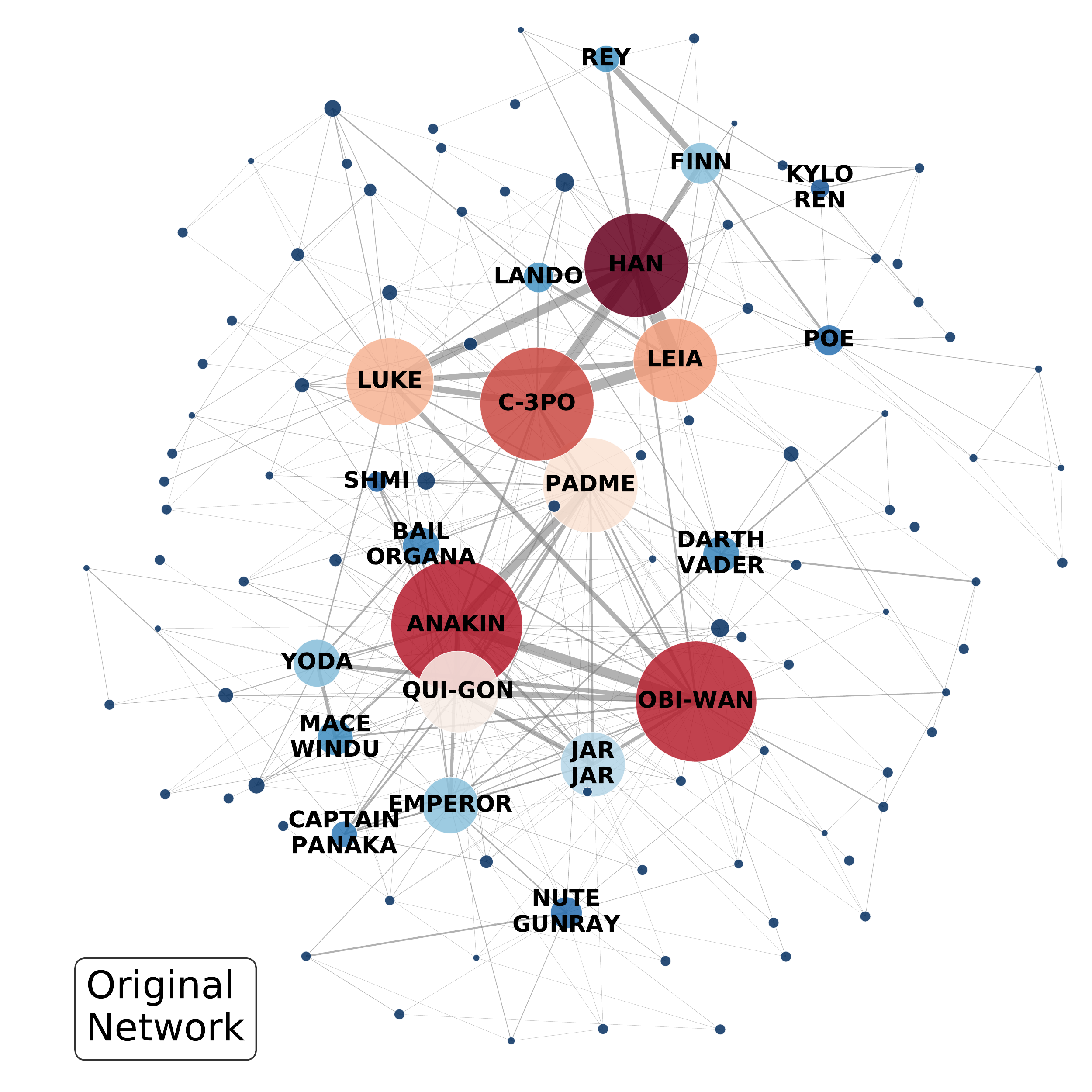}
\includegraphics[width=0.48\columnwidth]{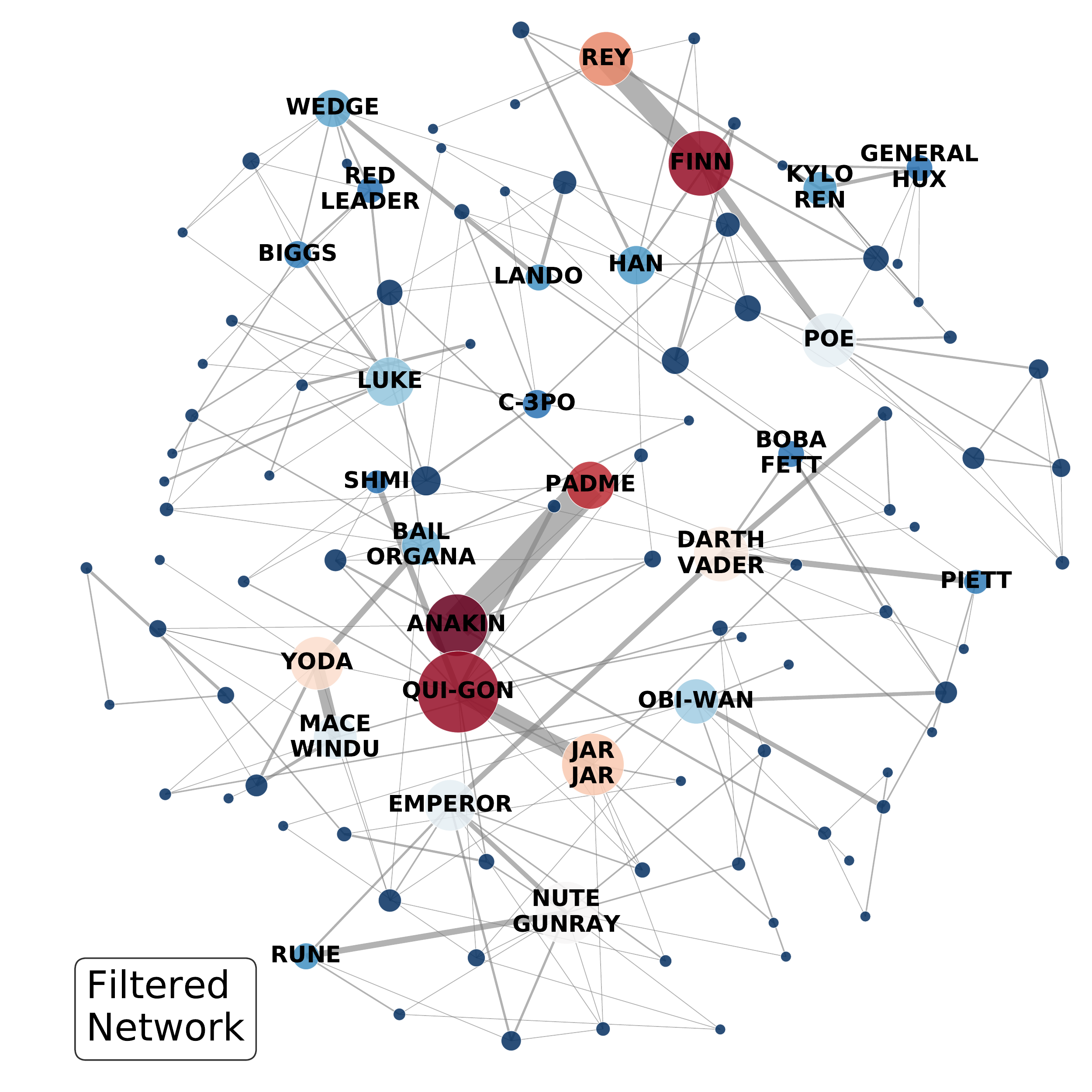}
\caption{Original graph showing the interactions among Star Wars characters (left) and corresponding pruned graph (right), according to the local ECM filter with $p$-value equal to $\widetilde{\gamma} = 0.05$ (bottom). The size of the nodes are proportional to the products of their degrees and strengths. We highlight also the most important connections.}

\label{fig:starwars_filtered}
\end{figure*}

We conclude our portfolio of datasets with the Star Wars movie saga one. The effects of filter aggressivity on topological quantities shown in Fig.~\ref{fig:starwars_filter_pval} are in line with similar results for other datasets. However the amount of retained information for $\widetilde{\gamma} = 0.05$ is much higher than any other case. This is probably due to the fact that these networks are already very sparse and therefore the statistical significance of their links is high. Considering the full interactions dataset, the visual inspection (Fig.~\ref{fig:starwars_filtered}) of the original and filtered networks permits to identify those characters playing a key role. In particular, the centrality of Darth Vader, Luke Skywalker and Obi-Wan Kenobi clearly increases while for other characters like C3P0 and Jar Jar Binks this is the opposite, showing once more the usefulness of the ECM filter.

%
%
%
%

%
%
%

\end{document}